\documentclass[journal=jctcce,manuscript=article]{achemso}
\usepackage[T1]{fontenc}
\usepackage[utf8]{inputenc}
\usepackage{lmodern}
\usepackage{amsmath,amssymb,bm,mathrsfs}
\usepackage{algorithm}
\usepackage{algpseudocode}
\usepackage{graphicx,booktabs,tabularx,array}
\usepackage{xcolor}
\usepackage{microtype}
\usepackage{tikz}
\usetikzlibrary{calc,shapes.geometric}
\usetikzlibrary{arrows.meta,positioning,fit,backgrounds}
\setkeys{acs}{articletitle=true,chaptertitle=true,doi=true,maxauthors=0}
\usepackage[hidelinks]{hyperref}
\hypersetup{pdftitle={ Efficient Spectra and Dynamics with Tangent-Space Chebyshev Tree Tensor Networks },pdfsubject={Manuscript for Journal of Chemical Theory and Computation}}
\SectionNumbersOn
\definecolor{tacheblue}{RGB}{31,79,110}
\definecolor{tachegreen}{RGB}{38,111,90}

\title{Efficient Spectra and Dynamics with Tangent-Space Chebyshev Tree Tensor Networks}
\author{Zhangyu Yang}
\affiliation{Guangdong Basic Research Center of Excellence for Aggregate Science, School of Science and Engineering, The Chinese University of Hong Kong (Shenzhen), Longgang, Shenzhen 518172, Guangdong, People's Republic of China}
\author{Xingyao Wu}
\affiliation{Technical University of Munich, CIT, Department of Computer Science, Boltzmannstra{\ss}e 3, 85748 Garching, Germany}
\author{Shuo Sun}
\affiliation{Technical University of Munich, CIT, Department of Computer Science, Boltzmannstra{\ss}e 3, 85748 Garching, Germany}
\author{Yu Wang}
\email{ywang96@sdu.edu.cn}
\affiliation{Key Laboratory for Colloid and Interface Chemistry, Ministry of Education, School of Chemistry and Chemical Engineering, Shandong University, Qingdao 266237, People's Republic of China}
\author{Weitang Li}
\email{liwt31@gmail.com }
\affiliation{Guangdong Basic Research Center of Excellence for Aggregate Science, School of Science and Engineering, The Chinese University of Hong Kong (Shenzhen), Longgang, Shenzhen 518172, Guangdong, People's Republic of China}

\author{Jan von Delft}
\affiliation{Arnold Sommerfeld Center for Theoretical Physics, Center for NanoScience, and Munich Center for Quantum Science and Technology, Ludwig-Maximilians-Universit\"at M\"unchen, 80333 Munich, Germany}
\author{Christian B. Mendl}
\affiliation{Technical University of Munich, CIT, Department of Computer Science, Boltzmannstra{\ss}e 3, 85748 Garching, Germany}
\alsoaffiliation{Technical University of Munich, Institute for Advanced Study, Lichtenbergstra{\ss}e 2a, 85748 Garching, Germany}
\author{Zhigang Shuai}
\affiliation{Guangdong Basic Research Center of Excellence for Aggregate Science, School of Science and Engineering, The Chinese University of Hong Kong (Shenzhen), Longgang, Shenzhen 518172, Guangdong, People's Republic of China}
\alsoaffiliation{Key Lab of Organic Optoelectronics and Molecular Engineering of Ministry of Education, Department of Chemistry, Tsinghua University, Beijing 100084, People's Republic of China}

\AtBeginDocument{\singlespacing}
\begin{document}

\begin{abstract}
We present a tangent-space Chebyshev method (TaChe), which combines tree tensor network states with Chebyshev recursion and maximum entropy spectral analysis in a tree tensor network tangent space, for quantum many-body spectra and dynamics. Implemented tensor network state calculations in the tangent space of the reference state, the core advantage of TaChe is that it can substantially reduce the computational complexity, memory usage, and numerical instability associated with tensor network state compression. Implemented on the GPU platform, it produces accurate spectra with speedups of up to $46.8$ over the previous Chebyshev tensor network method and $13.7$ over one-site time-dependent variational principle propagation in the reported benchmarks. Applications to acetonitrile, thermal aggregate emission, finite-temperature charge transport on a \(6\times6\) lattice, and internal conversion in aggregates of up to 100 molecules demonstrate the efficiency and versatility of TaChe in simulating complex molecular systems.
\end{abstract}

\section{Introduction}
\label{sec:introduction}

Quantum many-body spectra and dynamical correlation functions connect quantum states to experimentally accessible observables. Their accurate calculation remains a central challenge in quantum chemistry, owing to the exponential growth of the many-body Hilbert space with system size. Tensor-network states exploit the entanglement structure of these states to reduce the cost of the many-body representation. Matrix product states (MPS), tree tensor network states (TTNS), and their associated density matrix renormalization group (DMRG) algorithms have consequently become useful tools for electronic structure and molecular quantum dynamics~\cite{White1992,Schollwoeck2011,ChanSharma2011,ChanEtAl2016,Nakatani2013,BaiardiReiher2020}.

A common approach to computing spectra is to propagate a quantum state in real time and Fourier transform the resulting time-correlation function. Time-dependent density matrix renormalization group (TD-DMRG) methods, including those based on the time-dependent variational principle (TDVP), have demonstrated broad applicability to electronic, vibrational, and vibronic Hamiltonians~\cite{WhiteFeiguin2004,DaleyEtAl2004,HaegemanEtAl2011,HaegemanEtAl2016,PaeckelEtAl2019,RoncaEtAl2017,RenEtAl2018,BaiardiReiher2019,Baiardi2021}.  However, accurate characterization of long-time dynamics and high-resolution spectral features remains computationally demanding. Frequency-domain methods address the same function through correction vectors, Krylov subspaces, or polynomial expansions~\cite{Jeckelmann2002,DargelEtAl2012,JiangEtAl2020,LeeZhaiChan2023}. Chebyshev expansions are particularly useful because a single sequence of moments can determine both spectral functions and time  correlations~\cite{WeisseEtAl2006,HolznerEtAl2011,WolfEtAl2014,WolfEtAl2015}. Re-orthogonalized Chebyshev methods further use the generated tensor-network vectors to construct a reduced Hamiltonian for spectral reconstruction~\cite{XieEtAl2018,JiangEtAl2021}.

In conventional tensor-network implementations, repeated Hamiltonian applications increase the bond dimension, necessitating compression after each iteration to maintain a manageable representation. This compression, often performed through singular-value decompositions or variational optimization sweeps, can become a major computational bottleneck. Moreover, truncation errors disrupt the recurrence relations of iterative eigensolvers, therefore often requiring reorthogonalization or restarting, which further increases computational and memory costs~\cite{JiangEtAl2020,WangEtAl2026,XieEtAl2018}. Tangent-space methods offer an alternative by representing variations in a linear space around a tensor-network reference, allowing the projected Hamiltonian to act directly on tangent-space coefficients without repeated compression. This framework underlies variational excitation methods, TDVP, and DMRG linear-response approaches~\cite{HaegemanEtAl2011,HaegemanEtAl2012,HaegemanEtAl2013Post,WoutersEtAl2013,DorandoEtAl2009,NakataniEtAl2014}, and was further developed through the orthogonal projector formalism of Gleis, Li, and von Delft~\cite{GleisEtAl2022}. More recently, tangent-space Krylov (TaSK) calculations exploited this strategy to compute MPS spectral functions using Lanczos iteration~\cite{KovalskaEtAl2026}.  These advances are particularly attractive for tree tensor network states (TTNS), where the branching structure makes conventional variational compression more expensive than in MPS.

Here, we introduce the tangent-space Chebyshev method (TaChe) and formulate it within the TTNS representation. Unlike Lanczos methods used in TaSK, Chebyshev recursion is numerically more stable and does not suffer from the accumulation of roundoff errors, thereby eliminating the need for computationally expensive reorthogonalization procedures~\cite{SilverRoder1997}. Tree tensor networks allow strongly coupled degrees of freedom to be grouped into branches, naturally accommodating the molecular and vibrational structure of quantum-chemical Hamiltonians~\cite{Nakatani2013,Gunst2018,Larsson2019}. We construct both one-site and two-site tangent spaces and employ their projected Hamiltonians to generate scalar Chebyshev moments. By exploiting moment doubling, two Chebyshev moments are obtained per Hamiltonian application, while the three-term recurrence requires only a fixed number of tangent-vector buffers. Combined with GPU-accelerated tensor contractions and the reuse of reference environments, this approach avoids repeated tensor-network compression and reduces both computational overhead and memory requirements for spectral calculations.

We benchmark TaChe in four quantum-chemical applications. First, we compute the infrared spectrum of acetonitrile and the finite-temperature emission spectrum of a Frenkel--Holstein aggregate, comparing spectral accuracy and computational cost against reorthogonalization-based Chebyshev and TDVP methods at zero and finite temperatures, respectively. We further examine the advantages of the two-site tangent space by introducing a strongly coupled low-frequency vibrational mode. To investigate charge-carrier dynamics~\cite{Li2021ChargeTransport}, we calculate current--current correlation functions and optical conductivities on a $6\times6$ Holstein--Peierls lattice. Finally, we investigate the effects of excitonic coupling and its quantum fluctuations on nonradiative decay~\cite{Wang2023Nonradiative} through internal-conversion calculations on $6\times6$ and $10\times10$ aggregates. 

\section{Theoretical Background}
\label{sec:background}

\subsection{Chebyshev Representation of Spectra and Correlations}
\label{sec:chebyshev}

Spectral functions and time correlations can be evaluated from a common set of Chebyshev moments generated by repeated Hamiltonian applications~\cite{WeisseEtAl2006,HolznerEtAl2011,WolfEtAl2015}. Previous Chebyshev--MPS approaches such as ReCheMPS by Xie et al.~\cite{XieEtAl2018} and coCheb by Jiang et al.~\cite{JiangEtAl2021} employed orthogonalized Chebyshev vectors to construct an effective Hamiltonian in a Krylov subspace, effectively recasting spectral reconstruction as a
Krylov-subspace problem. The canonical-orthogonalization (coCheb) is described in the Supporting Information. 

Here, we instead use the Chebyshev moments directly for time-dependent correlations and spectral function. We first consider the zero-temperature case. For a normalized ground state $|\Psi_0\rangle$ satisfying
$H|\Psi_0\rangle=E_0|\Psi_0\rangle$, the spectral function and its corresponding correlation function are defined as
\begin{align}
F_{AB}(E;0)
&=
\langle\Psi_0|
A\,\delta\!\left(E-(H-E_0)\right)B
|\Psi_0\rangle,
\\
C_{AB}(t;0)
&=
\langle\Psi_0|
A e^{-i(H-E_0)t/\hbar}B
|\Psi_0\rangle
\nonumber\\
&=
\int_{-\infty}^{\infty}
dE\,e^{-iEt/\hbar}F_{AB}(E;0).
\end{align}

For numerical stability of Chebyshev, the Hamiltonian and energy are
shifted and rescaled according to
\begin{equation}
\overline H=\frac{H-E_0-c}{a},
\qquad
x=\frac{E-c}{a},
\end{equation}
where
\begin{equation}
c=\frac{E_{\max}+E_{\min}}{2},
\qquad
a=(1+\epsilon)\frac{E_{\max}-E_{\min}}{2},
\qquad \epsilon>0.
\end{equation}
Here $[E_{\min},E_{\max}]$ bounds the relevant transition
energies, and the small safety factor $\epsilon$ ensures
that the rescaled spectrum lies strictly within $(-1,1)$
on the relevant invariant space
\cite{WeisseEtAl2006,HolznerEtAl2011}.

Starting from $|t_0\rangle=B|\Psi_0\rangle$, the Chebyshev
vectors are generated through the three-term recursion
\begin{align}
|t_0\rangle &= B|\Psi_0\rangle,\\
|t_1\rangle &= \overline H|t_0\rangle,\\
|t_{n+1}\rangle
&=2\overline H|t_n\rangle-|t_{n-1}\rangle.
\end{align}
Consequently,
$|t_n\rangle=T_n(\overline H)B|\Psi_0\rangle$,
where $T_n(x)$ denotes the Chebyshev polynomial of the
first kind. Each recursion step requires only one
Hamiltonian application, and the corresponding moment is
\begin{equation}
\mu_n=
\langle\Psi_0|A|t_n\rangle
=
\langle\Psi_0|
A T_n(\overline H)B
|\Psi_0\rangle.
\label{eq:chebyshev_moment}
\end{equation}

The Chebyshev moments can directly determine the time-dependent
correlation function through the Chebyshev--Bessel expansion
of the evolution operator
\cite{TalEzerKosloff1984,HolznerEtAl2011,WolfEtAl2015}.
Using the first $M$ moments, we obtain
\begin{equation}
C_{AB}^{(M)}(t;0)
=
e^{-ict/\hbar}
\left[
\mu_0J_0\left(\frac{at}{\hbar}\right)
+
2\sum_{n=1}^{M-1}
(-i)^n\mu_nJ_n\left(\frac{at}{\hbar}\right)
\right],
\label{eq:chebyshev_correlation}
\end{equation}
where $J_n$ is the Bessel function of the first kind.

At finite temperature, the practical calculations here are evaluated using purification, which represents a mixed state as a pure state in an enlarged physical-auxiliary Hilbert space\cite{RenEtAl2018,JiangEtAl2020}. For a Hamiltonian $H$, the thermal density operator is
$ \rho_\beta=\frac{e^{-\beta H}}{Z}, Z=\operatorname{Tr}e^{-\beta H}, \beta=(k_{\mathrm B}T)^{-1}.$
Introducing an auxiliary copy of the physical Hilbert space, the normalized infinite-temperature purification is
\begin{equation}
|\Phi_0\rangle=
\bigotimes_{p=1}^{N}
\left(
\frac{1}{\sqrt{d_p}}
\sum_{\sigma_p=1}^{d_p}
|\sigma_p\rangle_{\mathrm P}
|\sigma_p\rangle_{\mathrm A}
\right),
\label{eq:purification_initial}
\end{equation}
where $d_p$ is the local basis dimension.
Each factor is a maximally entangled physical--auxiliary pair, so tracing out the auxiliary space gives the normalized identity operator on the physical space. The finite-temperature purification is then obtained by imaginary-time evolution,
\begin{equation}
|\Phi_\beta\rangle=
\frac{
\left(e^{-\beta H/2}\otimes I_{\mathrm A}\right)|\Phi_0\rangle
}{
\left\|
\left(e^{-\beta H/2}\otimes I_{\mathrm A}\right)|\Phi_0\rangle
\right\|
},
\label{eq:chebyshev_purification}
\end{equation}
which satisfies
$\operatorname{Tr}_{\mathrm A}
|\Phi_\beta\rangle\langle\Phi_\beta|=\rho_\beta$.
In the TTNS representation, each physical coordinate and its
auxiliary partner are grouped at the same node.
Starting from Eq.~\ref{eq:purification_initial}, the physical
coordinates are propagated over the imaginary-time interval
$0\leq\tau\leq\beta/2$, while the auxiliary coordinates are
left unchanged.
Two-site TDVP updates allow the bond spaces to adapt during propagation, subject to a maximum bond dimension $D$, and the state is normalized after each step. For physical operators $A$ and $B$, we define $ |b_\beta\rangle=(B\otimes I_{\mathrm A})|\Phi_\beta\rangle,
|a_\beta\rangle=(A^\dagger\otimes I_{\mathrm A}) |\Phi_\beta\rangle.$
The corresponding Hermitian Liouvillian is
\begin{equation}
\mathcal L=
H\otimes I_{\mathrm A}
-
I_{\mathrm P}\otimes H^{\mathsf T},
\label{eq:chebyshev_liouvillian}
\end{equation}
where the transpose is taken in the basis defining the physical--auxiliary pairs.
Its eigenvalues are transition energies $E_f-E_i$. The finite-temperature spectral function and correlation function are therefore
\begin{align}
F_{AB}(E;T)
&=
\langle a_\beta|
\delta\!\left(E-\mathcal L\right)
|b_\beta\rangle,
\\
C_{AB}(t;T)
&=
\operatorname{Tr}\!\left[\rho_\beta A(t)B\right]
\nonumber\\
&=
\langle a_\beta|
e^{-i\mathcal L t/\hbar}
|b_\beta\rangle
\nonumber\\
&=
\int_{-\infty}^{\infty}
dE\,e^{-iEt/\hbar}F_{AB}(E;T),
\end{align}
where $A(t)=e^{iHt/\hbar}Ae^{-iHt/\hbar}$. 

The Chebyshev formulation can be straightforwardly extended to finite temperatures
by replacing the Hamiltonian $H$ with the Liouvillian $\mathcal{L}$.
The rescaled Liouvillian and the corresponding spectral variable are defined as
\begin{equation}
    \overline{\mathcal{L}} = \frac{\mathcal{L}-cI}{a},
    \qquad
    x = \frac{E-c}{a},
\end{equation}
where
\begin{equation}
    c = \frac{E_{\max}+E_{\min}}{2},
    \qquad
    a = (1+\epsilon)\frac{E_{\max}-E_{\min}}{2},
    \qquad \epsilon > 0.
\end{equation}
Here, $[E_{\min},E_{\max}]$ bounds the Liouvillian spectrum. Unlike the zero-temperature Hamiltonian, $\mathcal L$ already represents energy differences and requires no ground-state energy subtraction.
Starting from $|t_0^{(T)}\rangle=|b_\beta\rangle$, the finite-temperature Chebyshev vectors are generated again through the three-term recurrence relation. Consequently, $|t_n^{(T)}\rangle=T_n(\overline{\mathcal L})|b_\beta\rangle$, and the corresponding moment is
\begin{align}
\mu_n(T)
&=
\langle a_\beta|t_n^{(T)}\rangle
\nonumber\\
&=
\langle\Phi_\beta|
(A\otimes I_{\mathrm A})
T_n(\overline{\mathcal L})
(B\otimes I_{\mathrm A})
|\Phi_\beta\rangle
\nonumber.
\label{eq:chebyshev_moment_ft}
\end{align}

For non-negative spectral functions ($A=B^\dagger$), we can employ maximum-entropy reconstruction\cite{SilverRoder1997} using the Chebyshev moments. Maximum-entropy reconstruction may provide higher spectral resolution compared to the widely used kernel polynomial method (KPM) in the tensor networks Chebyshev. Moreover, performing the Chebyshev expansion in tangent space avoids the truncation errors, potentially improving the accuracy and reliability of maximum-entropy reconstruction. Specifically, we adopt Burg's spectral entropy-rate formulation\cite{Burg1975,ChoiCover1984}.
Suppressing the temperature argument, the normalized moments $r_n=\mu_n/\mu_0$, with $\mu_0>0$, are interpreted as cosine Fourier moments under
the transformation $x=\cos\theta$.
The resulting maximum-entropy angular density is
\begin{equation}
P_{\mathrm{ME}}(\theta)
=
\frac{v_p}{
\left|
1+\sum_{j=1}^{p}\alpha_j e^{-ij\theta}
\right|^2
},
\qquad p=M-1,
\end{equation}
where the coefficients $\alpha_j$ and normalization
$v_p$ are determined from the moments through the
Yule--Walker equations.
Transforming back to the energy domain gives
\begin{equation}
F_{\mathrm{ME}}(E)
=
\frac{
\mu_0 P_{\mathrm{ME}}(\arccos x)
}{
\pi a\sqrt{1-x^2}
},
\qquad x=\frac{E-c}{a}.
\label{eq:maximum_entropy_spectrum}
\end{equation}
The corresponding correlation function can also be
obtained by Fourier transforming the reconstructed
spectrum, although direct evaluation through
Eq.~\eqref{eq:chebyshev_correlation} avoids the
additional approximation introduced by spectral
reconstruction.

\subsection{One-Site and Two-Site TTNS Tangent Spaces}
\label{sec:tangent_space}

The most important step in the Chebyshev spectral method is applying Hamiltonian to a quantum state. We now describe the action of the Hamiltonian in tangent spaces
of a reference TTNS, adapting the  formalism~\cite{GleisEtAl2022,KovalskaEtAl2026} to a tree. Let $H$ be represented
as a tree tensor network operator (TTNO) on a tree with $N$ sites
and local Hilbert-space dimensions $d_v$. Its normalized
reference state $|\Psi_{ref}\rangle$ is a TTNS in canonical form with bond
dimensions $D_{uv}$, taken to be its nonzero Schmidt ranks.
With site $r$ chosen as the orthogonality center,
\begin{equation}
 |\Psi_{ref}\rangle
 = \begin{tikzpicture}[
    baseline={([yshift=-0.5ex]current bounding box.center)},
    scale=0.8,
    every node/.style={transform shape},
    ttn triangle/.style={
        regular polygon,
        regular polygon sides=3,
        draw=black,
        line width=0.7pt,
        minimum size=0.95cm,
        inner sep=0pt
    },
    ttn left/.style={
        ttn triangle,
        fill=ttnsblue,
        shape border rotate=90
    },
    ttn right/.style={
        ttn triangle,
        fill=ttnsblue,
        shape border rotate=270
    },
    ttn vertical/.style={
        ttn triangle,
        fill=ttnsblue,
        shape border rotate=0
    },
    ttn center/.style={
    circle,
    draw=black,
    fill=ttnsblue,
    line width=0.7pt,
    minimum size=0.85cm,
    inner sep=0pt,
    font=\small
},
    ttn bond/.style={
        draw=black,
        line width=0.75pt
    }
]
\definecolor{ttnsblue}{RGB}{80,135,220}
\definecolor{ttnsgray}{RGB}{190,190,190}

\node[ttn right]  (l2) at (0,0) {};
\node[ttn right]  (l1) at (1.35,0) {};
\node[ttn center] (c)  at (2.70,0) {$C_r$};
\node[ttn left]   (r1) at (4.05,0) {};
\node[ttn left]   (r2) at (5.40,0) {};

\node[ttn vertical] (d1) at (2.70,-1.45) {};
\node[ttn vertical] (d2) at (2.70,-2.90) {};

\draw[ttn bond] (l2) -- (l1);
\draw[ttn bond] (l1) -- (c);
\draw[ttn bond] (c) -- (r1);
\draw[ttn bond] (r1) -- (r2);
\draw[ttn bond] (c) -- (d1);
\draw[ttn bond] (d1) -- (d2);

\draw[ttn bond] (l2) -- ++(45:0.75cm);
\draw[ttn bond] (l1) -- ++(45:0.75cm);
\draw[ttn bond] (c)  -- ++(45:0.75cm);
\draw[ttn bond] (r1) -- ++(45:0.75cm);
\draw[ttn bond] (r2) -- ++(45:0.75cm);
\draw[ttn bond] (d1) -- ++(45:0.80cm);
\draw[ttn bond] (d2) -- ++(45:0.80cm);

\end{tikzpicture}
\,,
 \qquad \langle\Psi_{ref}|\Psi_{ref}\rangle=1.
 \label{eq:ttns_canonical}
\end{equation}
The center tensor is denoted by $C_r$. For each nonroot site $v$,
let $p^{(v)}$ be its parent. The tensor $A[v]$ is an isometry
toward $p^{(v)}$, satisfying $A[v]^\dagger A[v]=I$ when its physical
and child indices are grouped as rows and its parent index as
columns.  $A[v\to w]$ denotes the reference isometry at $v$
directed toward a neighboring site $w$, so that
$A[v]=A[v\to p^{(v)}]$.  For
$A[v\to w]$, the row index groups the physical index and all
incident virtual indices except the one toward $w$. 
We denote the complete orthogonal-complement isometries by
$\overline A[v\to w]$, with
$\overline A\,\overline A^\dagger=I-AA^\dagger$
in the corresponding local effective space.
Blue and gray triangles denote kept and complementary
isometries, respectively; the upper layer of a projector
diagram is complex conjugated.

Let $\mathcal F_v[T]$ denote the state obtained by inserting a
tensor $T$ at $v$ and contracting it with the orthonormal kept
bases on all incident branches. Thus
$\mathcal F_v^\dagger\mathcal F_v=I$ and
$\mathcal F_r[C_r]=|\Psi_{ref}\rangle$.
One-site variations span the tangent space $\mathbb V^{1s}$.
Removing the reference-state direction defines
$\mathbb V^{1\perp}
 =\mathbb V^{1s}\cap
  \operatorname{span}\{|\Psi_{ref}\rangle\}^{\perp}$,
whose projector has the orthogonal decomposition
\begin{equation}
 \mathcal P^{1\perp}
 = P_r+\sum_{v\ne r}
   \underbrace{
\begin{tikzpicture}[
    baseline={([yshift=-0.5ex]current bounding box.center)},
    scale=0.8,
    every node/.style={transform shape},
    proj triangle/.style={
        regular polygon,
        regular polygon sides=3,
        draw=black,
        line width=0.7pt,
        minimum size=0.95cm,
        inner sep=0pt
    },
    proj left/.style={
        proj triangle,
        shape border rotate=90
    },
    proj right/.style={
        proj triangle,
        shape border rotate=270
    },
    proj up/.style={
        proj triangle,
        shape border rotate=0
    },
    proj bond/.style={
        draw=black,
        line width=0.75pt
    },
    proj bridge/.style={
        proj bond,
        line join=round
    }
]
\definecolor{projblue}{RGB}{80,135,220}
\definecolor{projback}{RGB}{170,205,240}
\definecolor{projgray}{RGB}{190,190,190}

\coordinate (F-l2) at (0,0);
\coordinate (F-l1) at (1.35,0);
\coordinate (F-c)  at (2.70,0);
\coordinate (F-r1) at (4.05,0);
\coordinate (F-r2) at (5.40,0);
\coordinate (F-d1) at (2.70,-1.45);
\coordinate (F-d2) at (2.70,-2.90);

\coordinate (F-cutL) at (1.90,0);
\coordinate (F-cutR) at (2.15,0);

\foreach \n in {l2,l1,c,r1,r2,d1,d2,cutL,cutR}{
    \coordinate (B-\n) at ($(F-\n)+(0.6,0.55)$);
}

\foreach \layer in {B,F}{
    \draw[proj bond] (\layer-l2) -- (\layer-l1);
    \draw[proj bond] (\layer-c) -- (\layer-r1);
    \draw[proj bond] (\layer-r1) -- (\layer-r2);
    \draw[proj bond] (\layer-c) -- (\layer-d1);
    \draw[proj bond] (\layer-d1) -- (\layer-d2);
}

\draw[proj bridge]
    (F-l1) -- (F-cutL) -- (B-cutL) -- (B-l1);

\draw[proj bridge]
    (F-c) -- (F-cutR) -- (B-cutR) -- (B-c);

\foreach \n in {l2,l1,c,r1,r2}{
    \draw[proj bond] (F-\n) -- ++(225:0.75cm);
    \draw[proj bond] (B-\n) -- ++(45:0.75cm);
}
\foreach \n in {d1,d2}{
    \draw[proj bond] (F-\n) -- ++(225:0.80cm);
    \draw[proj bond] (B-\n) -- ++(45:0.80cm);
}

\foreach \layer/\col in {B/projblue,F/projblue}{
    \node[proj right,fill=\col] at (\layer-l2) {};

    \node[proj right,fill=projgray,font=\small] at (\layer-l1) {D};

    \node[proj left,fill=\col,font=\small] at (\layer-c) {K};

    \node[proj left,fill=\col] at (\layer-r1) {};
    \node[proj left,fill=\col] at (\layer-r2) {};
    \node[proj up,fill=\col] at (\layer-d1) {};
    \node[proj up,fill=\col] at (\layer-d2) {};
}

\node[font=\small,anchor=north] at (1.05,-0.72) {$v$};
\node[font=\small,anchor=south] at (3.40,1.10) {$p(v)$};
\end{tikzpicture}
}_{P_v}.
 \label{eq:ttns_one_site_projector}
\end{equation}
Here $P_v=\mathcal F_v\mathcal Q_v\mathcal F_v^\dagger$, where
the local orthogonal projectors act as
\begin{equation}
 \mathcal Q_v Y =
 \begin{cases}
  Y-C_r\langle C_r,Y\rangle_{\mathrm F}, & v=r,\\[2pt]
  \bigl(I-A[v]A[v]^\dagger\bigr)Y, & v\ne r,
 \end{cases}
 \label{eq:ttns_local_projectors}
\end{equation}
and $\langle\cdot,\cdot\rangle_{\mathrm F}$ is the Frobenius
inner product. In particular, $P_r$ removes only the component
parallel to $C_r$ from the root variation space.

A vector in $\mathbb V^{1\perp}$ can therefore be written as
\begin{equation}
 |\Phi^{1\perp}(T)\rangle
 = \mathcal F_r[T[r]]
   +\sum_{v\ne r}
\begin{tikzpicture}[
    baseline={([yshift=-0.5ex]current bounding box.center)},
    scale=0.8,
    every node/.style={transform shape},
    tan triangle/.style={
        regular polygon,regular polygon sides=3,
        draw=black,line width=0.7pt,
        minimum size=0.95cm,inner sep=0pt,fill=tanblue
    },
    tan left/.style={tan triangle,shape border rotate=90},
    tan right/.style={tan triangle,shape border rotate=270},
    tan up/.style={tan triangle,shape border rotate=0},
    tan center/.style={
        circle,draw=black,fill=tangray,line width=0.7pt,
        minimum size=0.85cm,inner sep=0pt
    },
    tan bond/.style={draw=black,line width=0.75pt}
]
\definecolor{tanblue}{RGB}{80,135,220}
\definecolor{tangray}{RGB}{190,190,190}

\node[tan right]  (l2) at (0,0) {};
\node[tan center] (l1) at (1.35,0) {};
\node[tan left]   (c)  at (2.70,0) {};
\node[tan left]   (r1) at (4.05,0) {};
\node[tan left]   (r2) at (5.40,0) {};
\node[tan up] (d1) at (2.70,-1.45) {};
\node[tan up] (d2) at (2.70,-2.90) {};

\draw[tan bond] (l2) -- (l1);
\draw[tan bond] (l1) -- (c);
\draw[tan bond] (c) -- (r1);
\draw[tan bond] (r1) -- (r2);
\draw[tan bond] (c) -- (d1);
\draw[tan bond] (d1) -- (d2);

\foreach \n in {l2,l1,c,r1,r2}{
    \draw[tan bond] (\n) -- ++(45:0.75cm);
}
\foreach \n in {d1,d2}{
    \draw[tan bond] (\n) -- ++(45:0.80cm);
}

\node[font=\large,anchor=south] at (1.35,0.50) {$T[v]$};
\end{tikzpicture}
\,,
 \label{eq:ttns_one_site_vector}
\end{equation}
where the $v$th summand is $\mathcal F_v[T[v]]$ and the local
tensors satisfy
\begin{equation}
 A[v]^\dagger T[v]=0 \quad (v\ne r),
 \qquad
 \langle C_r,T[r]\rangle_{\mathrm F}=0.
 \label{eq:ttns_gauge}
\end{equation}
We have fixed the tangent-space gauge and made the
summands mutually orthogonal and orthogonal to $|\Psi_{ref}\rangle$.
The gray circle denotes the gauge-constrained variation tensor
$T[v]$.

Let $\mathbb V^{2s}$ be the span of variations of neighboring
two-site blocks. The additional directions beyond the one-site
space form
$\mathbb V^{2\perp}
 =\mathbb V^{2s}\cap(\mathbb V^{1s})^\perp$.
For each bond $\langle uv\rangle$, they are obtained from the
complementary sectors of $A[u\to v]$ and $A[v\to u]$, with all
remaining branches retained. Let $\mathcal G_{uv}$ denote the
isometric embedding that inserts a coefficient matrix $X^{uv}$
between these two complementary bases. The projector
$P_{uv}^{\mathrm{DD}}=\mathcal G_{uv}\mathcal G_{uv}^\dagger$
then gives
\begin{equation}
 \mathcal P^{2\perp}
 = \sum_{\langle uv\rangle}
   \underbrace{
\begin{tikzpicture}[
    baseline={([yshift=-0.5ex]current bounding box.center)},
    scale=0.8,
    every node/.style={transform shape},
    proj triangle/.style={
        regular polygon,
        regular polygon sides=3,
        draw=black,
        line width=0.7pt,
        minimum size=0.95cm,
        inner sep=0pt
    },
    proj left/.style={
        proj triangle,
        shape border rotate=90
    },
    proj right/.style={
        proj triangle,
        shape border rotate=270
    },
    proj up/.style={
        proj triangle,
        shape border rotate=0
    },
    proj bond/.style={
        draw=black,
        line width=0.75pt
    },
    proj bridge/.style={
        proj bond,
        line join=round
    }
]
\definecolor{projblue}{RGB}{80,135,220}
\definecolor{projback}{RGB}{170,205,240}
\definecolor{projgray}{RGB}{190,190,190}

\coordinate (F-l2) at (0,0);
\coordinate (F-l1) at (1.35,0);
\coordinate (F-c)  at (2.70,0);
\coordinate (F-r1) at (4.05,0);
\coordinate (F-r2) at (5.40,0);
\coordinate (F-d1) at (2.70,-1.45);
\coordinate (F-d2) at (2.70,-2.90);

\coordinate (F-cutL) at (1.90,0);
\coordinate (F-cutR) at (2.15,0);

\foreach \n in {l2,l1,c,r1,r2,d1,d2,cutL,cutR}{
    \coordinate (B-\n) at ($(F-\n)+(0.6,0.55)$);
}

\foreach \layer in {B,F}{
    \draw[proj bond] (\layer-l2) -- (\layer-l1);
    \draw[proj bond] (\layer-c) -- (\layer-r1);
    \draw[proj bond] (\layer-r1) -- (\layer-r2);
    \draw[proj bond] (\layer-c) -- (\layer-d1);
    \draw[proj bond] (\layer-d1) -- (\layer-d2);
}

\draw[proj bridge]
    (F-l1) -- (F-cutL) -- (B-cutL) -- (B-l1);

\draw[proj bridge]
    (F-c) -- (F-cutR) -- (B-cutR) -- (B-c);

\foreach \n in {l2,l1,c,r1,r2}{
    \draw[proj bond] (F-\n) -- ++(225:0.75cm);
    \draw[proj bond] (B-\n) -- ++(45:0.75cm);
}
\foreach \n in {d1,d2}{
    \draw[proj bond] (F-\n) -- ++(225:0.80cm);
    \draw[proj bond] (B-\n) -- ++(45:0.80cm);
}

\foreach \layer/\col in {B/projblue,F/projblue}{
    \node[proj right,fill=\col] at (\layer-l2) {};

    \node[proj right,fill=projgray,font=\small] at (\layer-l1) {D};

    \node[proj left,fill=projgray,font=\small] at (\layer-c) {D};

    \node[proj left,fill=\col] at (\layer-r1) {};
    \node[proj left,fill=\col] at (\layer-r2) {};
    \node[proj up,fill=\col] at (\layer-d1) {};
    \node[proj up,fill=\col] at (\layer-d2) {};
}

\node[font=\small,anchor=north] at (1.05,-0.72) {$u$};
\node[font=\small,anchor=south] at (3.40,1.10) {$v$};
\end{tikzpicture}
}
              _{P_{uv}^{\mathrm{DD}}},
 \label{eq:ttns_two_site_projector}
\end{equation}
where each bond is counted once. Components with a kept sector
at either endpoint already belong to $\mathbb V^{1s}$.
The DD sectors are mutually orthogonal and orthogonal to
$\mathbb V^{1s}$, and their vectors have the form
\begin{equation}
 |\Phi^{2\perp}(X)\rangle
 = \sum_{\langle uv\rangle}
\begin{tikzpicture}[
 baseline={([yshift=-0.5ex]current bounding box.center)},scale=0.8,
 every node/.style={transform shape},
 tri/.style={regular polygon,regular polygon sides=3,draw=black,
 line width=0.7pt,minimum size=0.95cm,inner sep=0pt,fill=twoblue},
 left/.style={tri,shape border rotate=90},
 right/.style={tri,shape border rotate=270},
 up/.style={tri,shape border rotate=0},
 bond/.style={draw=black,line width=0.75pt}]
\definecolor{twoblue}{RGB}{80,135,220}
\definecolor{twogray}{RGB}{190,190,190}
\node[right] (l) at (0,0) {};
\node[right,fill=twogray,font=\small] (u) at (1.35,0) {D};
\node[rectangle,draw=black,fill=twogray,minimum width=0.85cm,
 minimum height=0.65cm,inner sep=1pt] (X) at (2.65,0) {$X^{uv}$};
\node[left,fill=twogray,font=\small] (v) at (3.95,0) {D};
\node[left] (r1) at (5.30,0) {};
\node[left] (r2) at (6.65,0) {};
\node[up] (d1) at (3.95,-1.45) {};
\node[up] (d2) at (3.95,-2.90) {};
\draw[bond] (l)--(u)--(X)--(v)--(r1)--(r2);
\draw[bond] (v)--(d1)--(d2);
\foreach \n in {l,u,v,r1,r2,d1,d2}{
 \draw[bond] (\n)--++(45:0.75cm);
}
\node[anchor=north,font=\small] at (1.35,-0.55) {$u$};
\node[anchor=south,font=\small] at (3.65,0.55) {$v$};
\end{tikzpicture}
\,,
 \label{eq:ttns_two_site_vector}
\end{equation}
with each summand equal to $\mathcal G_{uv}[X^{uv}]$.
The enlarged variation space is therefore
$\mathbb V^{1\perp}\oplus\mathbb V^{2\perp}$.

For either choice of variation space, we define
\begin{equation}
 H_{\mathcal P}
 =\mathcal P(H-E_0)\mathcal P,
 \qquad
 E_0=\langle\Psi_{ref}|H|\Psi_{ref}\rangle,
 \label{eq:ttns_projected_hamiltonian}
\end{equation}
where $\mathcal P=\mathcal P^{1\perp}$ for the one-site space
and $\mathcal P=\mathcal P^{1\perp}+\mathcal P^{2\perp}$ for the
enlarged space. Write an input vector as
$|\Phi_i\rangle
 =|\Phi^{1\perp}(T_i)\rangle+|\Phi^{2\perp}(X_i)\rangle$,
with the second term omitted in the one-site calculation.
Since $\mathcal P|\Phi_i\rangle=|\Phi_i\rangle$, the tensors
representing
$|\widetilde\Phi_{i+1}\rangle=H_{\mathcal P}|\Phi_i\rangle$
are
\begin{equation}
 \begin{aligned}
  \widetilde T_{i+1}[v]
   &=\mathcal Q_v\!\left[
      \mathcal F_v^\dagger H|\Phi_i\rangle
     \right]-E_0T_i[v],\\
  \widetilde X_{i+1}^{uv}
   &=\mathcal G_{uv}^\dagger H|\Phi_i\rangle
     -E_0X_i^{uv}.
 \end{aligned}
 \label{eq:ttns_projected_action}
\end{equation}
Operationally, we contract the TTNO with the complete input
sum, leaving the output site or bond indices open, and project
onto the corresponding local sector. Following the tangent MPS
contraction scheme~\cite{KovalskaEtAl2026}, the sums are evaluated
recursively using ordinary reference-state environments together
with input-dependent environments that accumulate tangent
insertions along the branches, and therefore can be efficiently accelerated using GPU. Since computationally expensive truncation is not necessary, applying Hamiltonian in the tangent space is substantially faster compared to the regular Hilbert space.

Finally, linear combinations of vectors expressed in these
same fixed bases are formed by componentwise addition,
$T[v]\leftarrow aT[v]+bS[v]$ and
$X^{uv}\leftarrow aX^{uv}+bY^{uv}$, without changing the reference
bond dimensions.

\subsection{Tangent-Space Chebyshev Method}
\label{sec:tache}

TaChe evaluates Chebyshev moments within a TTNS tangent space. At zero temperature, the reference state is chosen to be the optimized ground-state TTNS \(|\Psi_0\rangle\), with energy \(E_0\) and bond dimension \(D\), for simplicity. The tangent space is defined by the orthogonal projector
\begin{equation}
\mathcal P=
\begin{cases}
\mathcal P^{1\perp}, & \text{one-site TaChe},\\
\mathcal P^{1\perp}+\mathcal P^{2\perp},
& \text{two-site TaChe}.
\end{cases}
\label{eq:tache_projector}
\end{equation}
Here $\mathcal P^{1\perp}$ projects onto one-site
variations orthogonal to the reference, while
$\mathcal P^{2\perp}$ selects additional two-site
variations orthogonal to the complete one-site
tangent space.

For the autocorrelation function with $A=B^\dagger$, we define the projected initial vector and Hamiltonian as introduced previously:
$|t_0\rangle=\mathcal P B|\Psi_0\rangle,
H_{\mathcal P}=\mathcal P(H-E_0)\mathcal P.$

The shift $c$ and scale $a$ are chosen as in the
preceding subsection, using bounds on the spectrum
of $H_{\mathcal P}$ in the retained tangent space.
The rescaled projected Hamiltonian is
\begin{equation}
\overline H_{\mathcal P}
=
\frac{H_{\mathcal P}-c\mathcal P}{a},
\label{eq:tache_rescaled}
\end{equation}
whose spectrum lies strictly within $(-1,1)$ on
the retained space.

The Chebyshev vectors are generated through the
three-term recursion
\begin{align}
|t_1\rangle
&=\overline H_{\mathcal P}|t_0\rangle,\\
|t_{n+1}\rangle
&=2\overline H_{\mathcal P}|t_n\rangle
-|t_{n-1}\rangle,
\end{align}
and the corresponding moments are
\begin{equation}
\mu_n
=
\langle t_0|t_n\rangle
=
\langle t_0|
T_n(\overline H_{\mathcal P})
|t_0\rangle.
\label{eq:tache_moments}
\end{equation}
The zeroth moment
$\mu_0=\langle t_0|t_0\rangle$ gives the spectral
weight retained in the projected tangent space.
Equivalently, the initial vector can be normalized
once and the resulting moments multiplied by $\mu_0$.

For autocorrelation functions, Hermiticity of
$\overline H_{\mathcal P}$ and the Chebyshev product
identities allow additional moments to be obtained
without further Hamiltonian applications:
\begin{align}
\mu_{2n}
&=2\langle t_n|t_n\rangle-\mu_0,
\qquad n\geq1,\\
\mu_{2n+1}
&=2\langle t_n|t_{n+1}\rangle-\mu_1,
\qquad n\geq1.
\label{eq:tache_doubling}
\end{align}
Thus, $K$ applications of the projected Hamiltonian
provide $2K$ moments $\mu_0,\ldots,\mu_{2K-1}$
for reconstruction. The additional even moment
$\mu_{2K}$ can also be obtained from $|t_K\rangle$
without another Hamiltonian application.

For spectral computation, we reconstruct the spectral density from the projected moments by maximum entropy. Time correlations follow directly from the same moments through the Chebyshev--Bessel expansion. The overall TaChe procedure is summarized in Algorithm~\ref{alg:tache}.

\begin{algorithm}[t]
\caption{TaChe for spectral and time-dependent correlations}
\label{alg:tache}
\begin{algorithmic}[1]
\Require $H$, $|\Psi_0\rangle$, $E_0$, $B$,
projector $\mathcal P$, action count $K\geq1$,
requested observable
\State Construct fixed tangent frames and choose
rescaling parameters $(a,c)$
\State Define the projected Hamiltonian action
$\mathcal A(v)=\overline H_{\mathcal P}v$
\State $|t_0\rangle\gets
\mathcal P B|\Psi_0\rangle$
\State $\mu_0\gets\langle t_0|t_0\rangle$
\If{$\mu_0=0$}
  \State \Return Any elastic contribution;
  projected inelastic tangent vanishes
\EndIf
\State $|t_1\rangle\gets\mathcal A(|t_0\rangle)$
\State $\mu_1\gets\langle t_0|t_1\rangle$
\For{$n=1,\ldots,K-1$}
  \State $\mu_{2n}\gets
  2\langle t_n|t_n\rangle-\mu_0$
  \State $|t_{n+1}\rangle\gets
  2\mathcal A(|t_n\rangle)-|t_{n-1}\rangle$
  \State $\mu_{2n+1}\gets
  2\langle t_n|t_{n+1}\rangle-\mu_1$
  \State Discard $|t_{n-1}\rangle$
\EndFor
\State $M\gets 2K$
\If{the requested observable is $C(t)$}
  \State Evaluate Eq.~\eqref{eq:chebyshev_correlation}
  at the requested times
\ElsIf{the requested observable is an optical conductivity}
  \State Evaluate the broadened Chebyshev resolvent and conductivity
\Else
  \State Reconstruct $F_{\mathrm{ME}}^{\mathcal P}(E)$
  from $\{\mu_n\}_{n=0}^{M-1}$
  using maximum entropy
  \State Include the elastic contribution when
  required and apply any prescribed broadening
\EndIf
\end{algorithmic}
\end{algorithm}

At finite temperature, we use the thermal purification $|\Phi_\beta\rangle$
and the Hermitian Liouvillian $\mathcal{L}$ defined in Sec.~\ref{sec:chebyshev}
For a physical operator $B$, we have
$|b_\beta\rangle=(B\otimes I_{\mathrm A})|\Phi_\beta\rangle$.
We construct the one-site or two-site tangent space from the  TTNS frames of $|\Phi_\beta\rangle$. For each symmetry sector $s$ present in $|b_\beta\rangle$, let $P_{\beta,s}$ denote the orthogonal projector onto the response space in that sector. The projected source and generator are therefore
\[
  |t_{0,s}\rangle=P_{\beta,s}|b_\beta\rangle,
  \qquad
  \mathcal{L}_{\beta,s}=P_{\beta,s}\mathcal{L}P_{\beta,s}.
\]

At fixed physical and TTNO bond dimensions, each Chebyshev iteration requires only a single projected Hamiltonian application, scaling as $O(L\chi^4)$ for one-site TaChe, where $\chi$ is the TTNS bond dimension (for an MPS, a one-site TaChe iteration sweep cost \(O(L\chi^3)\) instead). Although one-site TDVP has the same leading scaling, it requires multiple local Hamiltonian applications to evaluate the exponential propagators. In contrast, conventional evaluation of $H|\psi\rangle$ involves repeated variational compression sweeps, often including two-site updates to achieve sufficient accuracy. Such updates are particularly expensive for TTNS, where two-site contractions scale as $O(L\chi^5)$ and internal-bond decompositions can cost $O(\chi^6)$ each.  The storage advantage follows from the Chebyshev recurrence. If the chosen tangent space has $m$ coefficients, moment doubling requires only the neighboring vectors $t_{n-1}$ and $t_n$ and a buffer for $t_{n+1}$. The recurrence-vector memory is therefore $O(m)$, independent of the number $M$ of retained moments; storing the moments adds $O(M)$ scalar memory. In comparison, coCheb retains its primitive tensor-network vectors and forms their overlap and Hamiltonian matrices, while a fully stored or reorthogonalized Lanczos basis requires storage that grows with the iteration count. A detailed analysis of the computational and memory costs is provided in the Supplemental Information.

\section{Results and Discussion}
\label{sec:results}

\subsection{Infrared Spectrum of Acetonitrile}
\label{sec:ch3cn}

\begin{figure}[tbp]
\centering
\includegraphics[width=\linewidth]{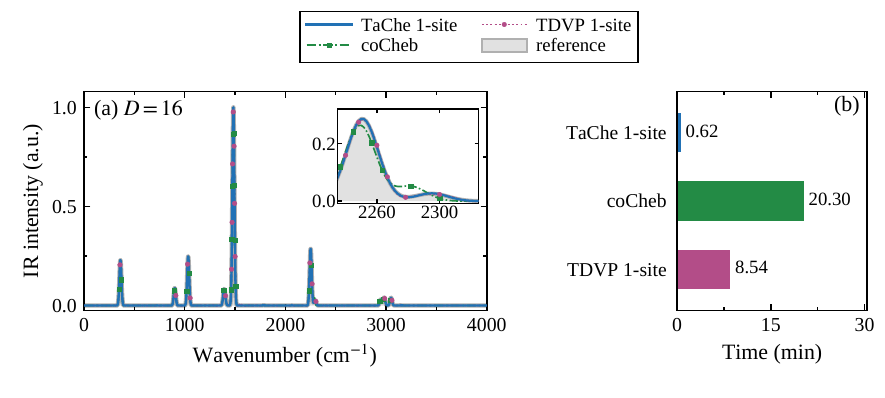}
\caption{Dipole spectral function of acetonitrile and computational cost.
(a) Spectra obtained with binary TTNS of bond dimension $D=16$,
using Gaussian broadening with $\sigma=10$~cm$^{-1}$.
One-site TaChe uses $250$ projected Hamiltonian
applications to obtain $500$ moments per direction; coCheb uses $500$ compressed
Chebyshev vectors per direction, with variational TTNS compression. One-site TDVP uses $\Delta t=320$ atomic units and $175$ steps.
The reference spectrum is obtained with two-site MPS TDVP of bond
dimension $D=50$, $\Delta t=100$ atomic units, and $1000$ time steps.
(b) GPU computation times summed over $x,y,z$, including
moment generation or time propagation and final spectral processing. Reference-state preparation and spectral-bound estimation are excluded.}
\label{fig:ch3cn}
\end{figure}

We first consider the zero-temperature dipole spectrum of acetonitrile,
a 12-mode benchmark for coupled anharmonic vibrations.
Previous vibrational DMRG studies have used this molecule to examine
local-basis and bond-dimension convergence~\cite{BaiardiStein2017}.
In dimensionless normal coordinates, the vibrational Hamiltonian is
\begin{align}
\hat H_{\mathrm{vib}}={}&
\frac12\sum_{k=1}^{12}\varepsilon_k
\left(-\frac{\partial^2}{\partial q_k^2}+q_k^2\right)
+\sum_{i\leq j\leq k}c_{ijk}q_iq_jq_k
\nonumber\\[-2pt]
&+\sum_{i\leq j\leq k\leq l}c_{ijkl}q_iq_jq_kq_l ,
\label{eq:ch3cn_hamiltonian}
\end{align}
where $\varepsilon_k$ are the harmonic mode energies and $c_{ijk}$
and $c_{ijkl}$ describe cubic and quartic anharmonic couplings.
We represent the vibrational state by a binary TTNS with bond dimension
$D=16$ and the mode hierarchy
\[
\bigl(((1,(5,6)),((7,8),(9,10))),((3,(2,4)),(11,12))\bigr).
\]
The ground state $|0\rangle$, obtained by two-site DMRG, defines the tangent space used by TaChe.

We use linear dipole operators
$\hat\mu_\alpha=\sum_k d_{\alpha k}q_k$, with $\alpha=x,y,z$.
The coefficients $d_{\alpha k}$ are taken from the transition-moment
output of a B3LYP/def2-TZVPD calculation performed with
ORCA~\cite{orca}.
The dipole spectral function is
\begin{equation}
A_{\mathrm{IR}}(E)=
\sum_{\alpha=x,y,z}
\langle0|\Delta\hat\mu_\alpha
\delta\!\left[E-(\hat H_{\mathrm{vib}}-E_0)\right]
\Delta\hat\mu_\alpha|0\rangle,
\label{eq:ch3cn_ir}
\end{equation}
where $E_0$ is the ground-state energy and
$\Delta\hat\mu_\alpha=\hat\mu_\alpha-
\langle0|\hat\mu_\alpha|0\rangle$.
The spectra in Fig.~\ref{fig:ch3cn} are broadened with a Gaussian
of standard deviation $\sigma=10$~cm$^{-1}$.

Here, we apply TaChe in the one-site tangent space.
Moment-doubling identities yield $500$ Chebyshev moments from 
$250$ Hamiltonian applications per Cartesian direction,
followed by maximum-entropy spectral reconstruction.
The coCheb comparison retains $500$ variationally compressed Chebyshev vectors per direction.

Figure~\ref{fig:ch3cn}(a) shows that one-site TaChe reproduces the positions and relative intensities of the reference bands across the displayed $0$--$4000$~cm$^{-1}$ range. The inset resolves a weak feature near $2270$~cm$^{-1}$, where TaChe follows the reference line shape and coCheb shows a larger deviation. Thus, a fixed one-site tangent space built from the $D=16$ ground state captures both the dominant transitions and weaker anharmonic spectral structure in this benchmark accurately.

This spectral agreement is obtained at substantially lower cost for TaChe. Summed over the three Cartesian directions, TaChe takes $37.4$~s on a single NVIDIA V100 GPU, compared with $20.3$~min for coCheb and $8.5$~min for one-site TDVP [Fig.~\ref{fig:ch3cn}(b)]. The corresponding speedups are $32.6$ and $13.7$. We note that these times include recurrence generation and final spectral processing; the coCheb time includes overlap and Hamiltonian matrix construction and the reduced eigenproblem. Ground-state preparation and spectral-bound estimation are excluded. 

\subsection{Aggregate Emission with Fast and Slow Vibrational Modes}
\label{sec:aggregate}

We next apply TaChe to finite-temperature emission from a
molecular aggregate. TD-DMRG and canonical-orthogonalized
Chebyshev (coCheb) methods have previously been applied to
the calculation of absorption and fluorescence spectra of
similar systems~\cite{RenEtAl2018,JiangEtAl2021}. We consider a Frenkel--Holstein model on an open chain of $L=10$ identical molecules. Following the setup of
Jiang et al.~\cite{JiangEtAl2021} and extending to next-nearest-neighbor electronic coupling, the Hamiltonian is written as
\begin{align}
\hat H_{\mathrm{FH}}={}&
\sum_{i=1}^{L}\epsilon_i\hat n_i
+\sum_{r=1}^{2}J_r\sum_{i=1}^{L-r}
\left(\hat a_i^\dagger\hat a_{i+r}
+\hat a_{i+r}^\dagger\hat a_i\right)
\nonumber\\
&+\sum_{i=1}^{L}\sum_\nu
\hbar\omega_\nu\hat b_{i\nu}^\dagger\hat b_{i\nu}
+\sum_{i=1}^{L}\sum_\nu
\hbar\omega_\nu g_\nu\hat n_i
\left(\hat b_{i\nu}^\dagger+\hat b_{i\nu}\right).
\label{eq:fh_extended}
\end{align}
Here $\hat a_i^\dagger$ creates a local electronic
excitation on molecule $i$, and
$\hat n_i=\hat a_i^\dagger\hat a_i$ is the corresponding
population operator. Each molecule is treated as a
two-level electronic system.
The operator $\hat b_{i\nu}^\dagger$ creates a vibrational
quantum in mode $\nu$ on molecule $i$, with angular
frequency $\omega_\nu$.
The dimensionless electron--vibration coupling $g_\nu$
is related to the Huang--Rhys factor by
$S_\nu=g_\nu^2$, with the displacement convention
$g_\nu=-\sqrt{S_\nu}$.
The electronic site energy $\epsilon_i$ is defined
in the undisplaced oscillator representation.
The nearest- and next-nearest-neighbor couplings are
$J_1/(hc)=700$~cm$^{-1}$ and
$J_2/(hc)=87.5$~cm$^{-1}$, respectively.

In the fast-only model, each molecule carries one
vibrational mode with
$\hbar\omega_f/(hc)=1400$~cm$^{-1}$ and $S_f=1$,
represented by five harmonic-oscillator basis functions.
The fast-plus-slow model additionally includes a mode
with $\hbar\omega_s/(hc)=140$~cm$^{-1}$ and $S_s=5$,
represented by twenty basis functions per molecule.
The corresponding reorganization energies are
$\lambda/(hc)=1400$~cm$^{-1}$ and
$2100$~cm$^{-1}$ for the fast-only and fast-plus-slow
models, respectively. All emission spectra are evaluated
at $T=298$~K, with the initial thermal ensemble
restricted to the one-exciton manifold,
$\sum_i\hat n_i=1$.

The finite-temperature state is prepared using the
purification formalism described in
Sec.~\ref{sec:chebyshev}. The purification is represented
by a binary-tree TTNS, with each physical coordinate
and its auxiliary copy placed at the same leaf.
Starting from the infinite-temperature purification,
we first apply
$I-\tau_0(H_e\otimes I_{\mathrm A})$,
with $\tau_0=10^{-6}$ a.u., to initialize the bond
sectors required by electronic hopping.
After normalization, the remaining imaginary-time
interval $\beta/2-\tau_0$ is evolved in $100$ equal
steps using two-site TDVP.
At each bond dimension $D$, TaChe, coCheb, and
real-time TDVP use the same prepared thermal state
$|\widetilde\Phi_\beta^{(D)}\rangle$.

\begin{figure}[tbp]
\centering
\includegraphics[width=5.9in]{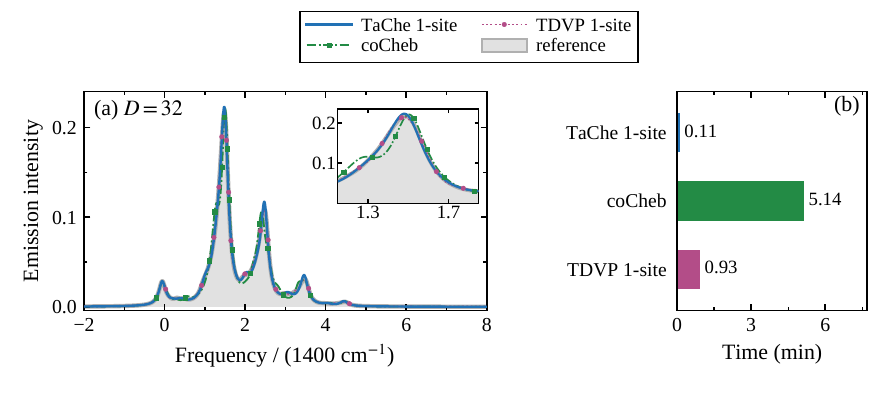}
\caption{Fast-mode aggregate emission and tangent time for the
$10$-molecule open chain at $298$~K.
(a) Binary-tree TTNS spectra at $D=32$, with
Lorentzian HWHM $140$~cm$^{-1}$. One-site TaChe uses $75$ projected-generator applications to obtain $150$ moments; coCheb retains $150$ compressed Chebyshev vectors.
The reference is two-site MPS TDVP with $D=64$, $\Delta t\,\omega_f=0.05$, and $t_{\max}\omega_f=100$.
The compared one-site TDVP uses $\Delta t\,\omega_f=0.5$ for $60$ steps, reaching $t_{\max}\omega_f=30$.
(b) GPU times under the same accounting as Figure~\ref{fig:ch3cn}.}
\label{fig:aggregate_fast}
\end{figure}

For the fast-mode aggregate, one-site TaChe computes the spectrum accurately [Fig.~\ref{fig:aggregate_fast}(a)]. Only $75$ projected-generator applications are needed to obtain the $150$ moments used for reconstruction. At the same reference bond dimension, coCheb uses $150$ compressed vectors and gives a less accurate line shape. One-site TDVP uses $\Delta t\,\omega_f=0.5$ for $60$ steps, reaching $t_{\max}\omega_f=30$. The tangent calculation takes only $6.6$~s on a single V100 GPU, compared with $308.2$~s for coCheb and $55.7$~s for one-site TDVP. TaChe is therefore $46.8$ and $8.5$ times faster, respectively, for the displayed spectra. Together with acetonitrile, this result shows that the fixed-space formulation is effective for both zero-temperature and finite-temperature computations. 

\begin{figure}[tbp]
\centering
\includegraphics[width=5.9in]{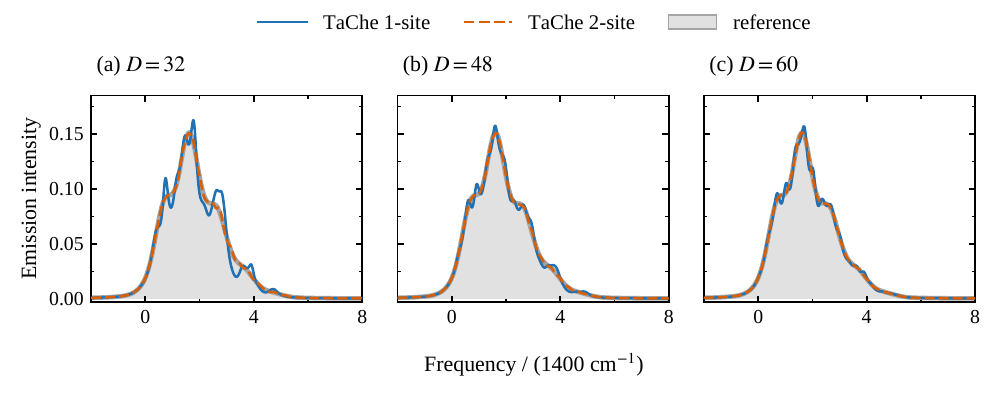}
\caption{Fast-and-slow aggregate emission for the
$10$-molecule open chain at $298$~K. (a) $D=32$, (b) $D=48$,
and (c) $D=60$. Both TaChe variants use $100$ projected-generator applications and $200$ moments. The reference is MPS TDVP with $D=64$, $\Delta t\,\omega_f=0.05$, and $t_{\max}\omega_f=100$.}
\label{fig:aggregate}
\end{figure}

The slow mode provides a more demanding test of the tangent space. At $298$~K, a $140$~cm$^{-1}$ vibration is thermally populated, and its large Huang--Rhys factor, $S_s=5$, produces a dense set of vibronic transitions~\cite{JiangEtAl2021,Spano2010}. Electronic hopping couples these local vibrational configurations across molecules. Figure~\ref{fig:aggregate} compares one-site and two-site TaChe at $D=32$, $48$, and $60$, using $100$ projected-generator applications and $200$ moments in each case. For this model, the two-site extension of TaChe substantially improves accuracy, suggesting that the one-site tangent space of the thermal reference state is too restrictive to adequately capture the relevant excitations.  We note that two-site TaChe can sometimes be more computationally expensive than one-site TDVP, however, it remains substantially more efficient and stable than the direct $H\psi$ approach employed in conventional frequency-domain tensor-network methods, thereby making it an attractive and computationally efficient alternative for frequency-domain calculations. 

\subsection{Charge Transport on a Two-Dimensional Molecular Lattice}
\label{sec:transport}

\begin{figure}[tbp]
\centering
\includegraphics[width=0.96\linewidth]{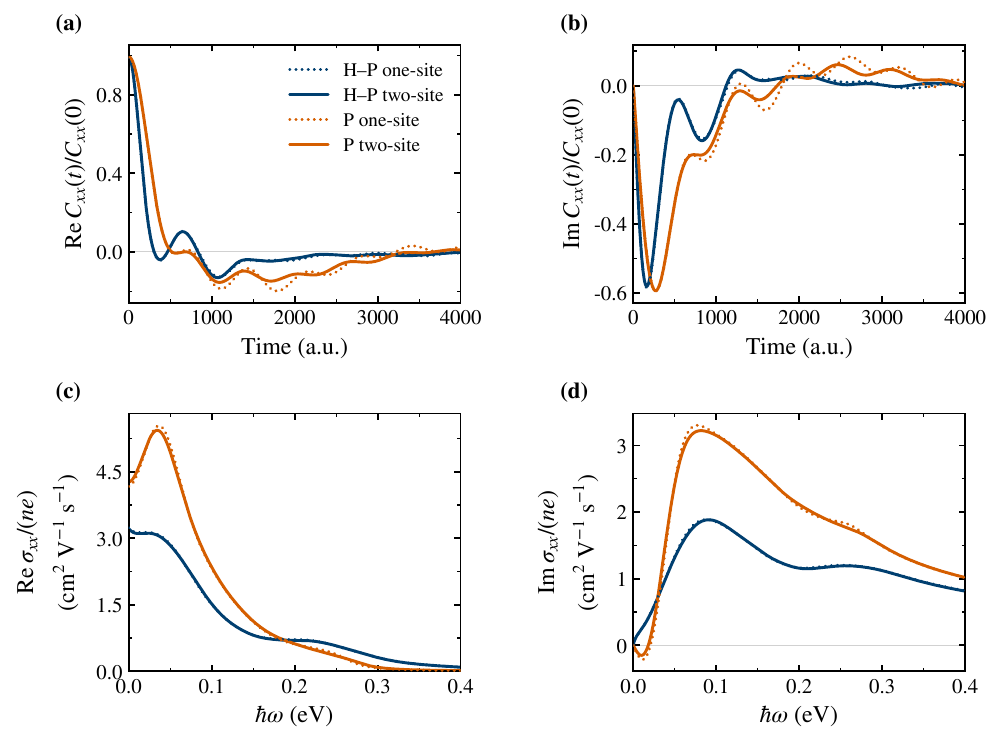}
\caption{Current correlations and optical conductivities for one
electron on a $6\times6$ open square lattice at $300$~K, with
$V_x=V_y=30$~meV and $\Delta V_x=\Delta V_y=20$~meV.
Blue and orange curves denote the Holstein--Peierls (H--P) and
pure Peierls (P) models, respectively. Dotted and solid curves show
one-site and two-site TaChe using the same $D=32$ thermal reference
for each model. (a,b) Real and imaginary parts of $C_{xx}(t)$,
normalized by the unprojected thermal-reference value
$\langle\hat j_x^2\rangle_\beta$.
(c,d) Real and imaginary parts of $\sigma_{xx}^{(\eta)}/(ne_0)$
at $\eta=20$~meV.}
\label{fig:transport}
\end{figure}

We now demonstrate TaChe's applicability to finite-temperature dynamics by investigating charge transport in
molecular semiconductors, where carrier motion is coupled to both local molecular reorganization and fluctuations in intermolecular
transfer integrals. The Holstein--Peierls model captures these two coupling mechanisms and provides a framework for understanding
phonon-assisted, transiently localized, band-like, and intermediate transport regimes.\cite{Li2021ChargeTransport}

Extending the previous study to 2D,\cite{Li2021ChargeTransport} we consider a two-dimensional $6\times6$ square lattice to examine
how local vibrations change the current response in the presence of fluctuating transfer integrals. The Hamiltonian is
\begin{align}
H={}&-\sum_{\alpha=x,y}V_\alpha
\sum_{\langle ij\rangle_\alpha}
(c_i^\dagger c_j+c_j^\dagger c_i)
\nonumber\\
&+\sum_{i,m}\epsilon_m b_{im}^\dagger b_{im}
+\epsilon_\theta\sum_{\langle ij\rangle}d_{ij}^\dagger d_{ij}
\nonumber\\
&+\sum_{i,m}g_m\epsilon_m
(b_{im}^\dagger+b_{im})n_i
\nonumber\\
&+\sum_{\alpha=x,y}g_{\theta,\alpha}\epsilon_\theta
\sum_{\langle ij\rangle_\alpha}
(d_{ij}^\dagger+d_{ij})
(c_i^\dagger c_j+c_j^\dagger c_i),
\label{eq:hp_lattice}
\end{align}
where $c_i^\dagger$ creates an electron on molecule $i$, $n_i=c_i^\dagger c_i$, $b_{im}^\dagger$ creates a quantum of local
intramolecular mode $m$, and $d_{ij}^\dagger$ creates a quantum of the intermolecular mode on bond $ij$. Each nearest-neighbor bond
is counted once and carries an independent oscillator; $\alpha=x,y$ labels horizontal and vertical bonds.
The parameters $\epsilon_m$ and $\epsilon_\theta$ are the corresponding vibrational quantum energies.

The Holstein interaction couples the carrier population to intramolecular displacements, whereas the Peierls interaction
modulates the hopping with bond displacements. The dimensionless coupling strengths are $g_m$ and $g_{\theta,\alpha}$, respectively.

We parameterize the Peierls coupling by the root-mean-square hopping fluctuation of an uncoupled thermal bond oscillator,
\begin{equation}
\Delta V_\alpha
=g_{\theta,\alpha}\epsilon_\theta
\sqrt{\coth\!\left(
\frac{\epsilon_\theta}{2k_{\mathrm B}T}
\right)}.
\label{eq:hp_fluctuation}
\end{equation}
We consider one electron at $300$~K with open boundaries, $V_x=V_y=30$~meV, and
$\Delta V_x=\Delta V_y=20$~meV at this temperature. The four local-mode parameters, bond-mode wavenumber of
$50$~cm$^{-1}$ ($\epsilon_\theta\simeq6.2$~meV), and lattice spacing of $7.2$~\AA\ are retained from the original
work.\cite{Li2021ChargeTransport} We compare the full Holstein--Peierls model (H--P), which includes
both couplings, with the pure Peierls model (P), obtained by setting $g_m=0$ for all local modes. The resulting decoupled local
oscillators are omitted in the P calculation.

The directional current correlation is
\begin{equation}
C_{\alpha\alpha}(t)
=\operatorname{Tr}[
\rho_\beta\hat j_\alpha(t)\hat j_\alpha(0)],
\qquad \alpha=x,y,
\label{eq:hp_directional_current}
\end{equation}
where $\rho_\beta$ is the normalized thermal density operator in the one-electron sector and $\hat j_\alpha$ is the charge current
in direction $\alpha$, including both the bare and phonon-modulated hopping contributions. The lattice and Hamiltonian are symmetric
under interchange of $x$ and $y$, so the exact correlations satisfy $C_{xx}(t)=C_{yy}(t)$. Finite TTNS approximations need not preserve
this equality exactly; Fig.~\ref{fig:transport} shows the $x$ result.

The correlated thermal purification $|\Phi_\beta\rangle$ is prepared by imaginary-time TTNS evolution with maximum bond
dimension $D=32$. The tree recursively partitions spatial groups of molecules, with branching subtrees for their electronic and
vibrational degrees of freedom and paired physical--auxiliary local bases.

We evaluate the complex current correlation $C_{xx}(t)$ directly
from Eq.~\eqref{eq:chebyshev_correlation} over $0$--$4000$ atomic units of time.
The same moments determine the optical conductivity normalized
by $ne_0$,
\begin{equation}
 \frac{\operatorname{Re}\sigma_{xx}^{(\eta)}(E/\hbar)}{n e_0}
 =\frac{1-e^{-\beta E}}{e_0 E}
 \operatorname{Re}\int_0^\infty
 e^{(iE-\eta)t/\hbar}C_{xx}^{\mathrm{TaChe}}(t)\,dt,
 \label{eq:hp_optical}
\end{equation}
where $E=\hbar\omega$, $\beta=(k_{\mathrm B}T)^{-1}$,
$n$ is the carrier density, and $e_0$ is the elementary charge.
The thermal prefactor is evaluated using its limit $\beta/e_0$ at $E=0$. We evaluate the integral directly through the Chebyshev
resolvent with $\eta=20$~meV; the displayed time window therefore does not truncate the optical reconstruction. The imaginary
conductivity is obtained by Kramers--Kronig completion of the even extension of the positive-frequency real conductivity.

For both one-site and two-site TaChe, we use $2048$ scalar Chebyshev moments for H--P and $1024$ for P. Moment doubling
generates these sequences from $1024$ and $512$ projected-Liouvillian applications, respectively, per direction.  
The two tangent spaces give the same qualitative distinction between the H--P and P models [Fig.~\ref{fig:transport}].
Including local molecular vibrations makes the negative minimum of the real correlation shallower. In the two-site results, the
normalized minimum changes from $-0.16$ for P to $-0.13$ for H--P. These changes illustrate how intramolecular reorganization modifies the negative current correlation associated with backscattering and redistributes the optical response, which also demonstrate TaChe's application to dynamics in finite-temperature.

\subsection{Internal Conversion with Fluctuating Excitonic Coupling}
\label{sec:ic}

Finally, we further benchmark TaChe by investigating internal conversion (IC), a nonradiative decay process governed by the molecular vibronic spectrum. Altering the two-mode aggregate model of Wang et al.~\cite{Wang2023Nonradiative}, we consider periodic $6\times6$ and $10\times10$ square lattices with quantum fluctuations in the nearest-neighbor excitonic coupling. The larger system, comprising $100$ molecules, provides a demanding test of the one-site TaChe formulation for spatially extended molecular aggregates. We examine how excitonic coupling and aggregate size influence the competition between molecular reorganization and excitation-energy reduction, demonstrating the applicability of TaChe to complex vibronic dynamics in extended chemical systems.

Each molecule has an IC-promoting mode $p$ and a Holstein-accepting mode $h$, both with energy $\epsilon_p/(hc)=\epsilon_h/(hc)=1400$~cm$^{-1}$.  The accepting mode has Huang--Rhys factor $S=0.5$,
giving the monomer reorganization energy $\lambda=S\epsilon_h=hc\times700$~cm$^{-1}$.
The monomer adiabatic excitation energy is $E_{\mathrm{ad}}/(hc)=14000$~cm$^{-1}$.
With $\hat a_i^{\dagger}$ creating a local excitation and $\hat n_i=\hat a_i^{\dagger}\hat a_i$, the two electronic sectors are
described by 
\begin{align}
\hat H_g={}&\sum_{i,\nu=p,h}\epsilon_\nu
\hat b_{i\nu}^{\dagger}\hat b_{i\nu}
+\epsilon_\theta\sum_{\langle ij\rangle}
\hat d_{ij}^{\dagger}\hat d_{ij},
\label{eq:ic_hg}\\
\hat H_e={}&\hat H_g+(E_{\mathrm{ad}}+\lambda)\sum_i\hat n_i
-\sqrt{S}\epsilon_h\sum_i\hat n_i
(\hat b_{ih}^{\dagger}+\hat b_{ih})
\nonumber\\
&+\sum_{\langle ij\rangle}
\left[J+\delta J(\hat d_{ij}^{\dagger}+\hat d_{ij})\right]
(\hat a_i^{\dagger}\hat a_j+\hat a_j^{\dagger}\hat a_i).
\label{eq:ic_he}
\end{align}
Common zero-point energies are omitted. The $\lambda$ counterterm
ensures that an isolated molecule has its relaxed excitation energy
at $E_{\mathrm{ad}}$. Each distinct nearest-neighbor bond carries an independent oscillator: there are $72$ such bonds on the $6\times6$ lattice and $200$ on the $10\times10$ lattice, with
$\epsilon_\theta/(hc)=50$~cm$^{-1}$.
At $T=0$, $\delta J$ equals the root-mean-square hopping fluctuation of an uncoupled bond oscillator. We compare
$\delta J=0$ and $\delta J/|J|=0.3$, with $J<0$ and
$|J|/\lambda=0,0.1,\ldots,1.0$.

Let $|\Psi_e\rangle$ be the normalized lowest state of $\hat H_e$
in the one-exciton sector, with energy $E_e$. The transition operator
can be written as $\hat B=\sum_i\hat a_i\hat\pi_{ip}$, where
$\hat\pi_{ip}=i(\hat b_{ip}^{\dagger}-\hat b_{ip})/\sqrt{2}$ is
the dimensionless promoting-mode momentum. The reverse transition
is $\hat A=\hat B^{\dagger}$. For a common nonadiabatic coupling
constant $v_{\mathrm{IC}}$, define
\begin{equation}
|\chi\rangle=\hat B|\Psi_e\rangle,\qquad
S_{\mathrm{IC}}(E)=\langle\chi|
\delta(E-\hat H_g)|\chi\rangle.
\label{eq:ic_spectral_function}
\end{equation}
The Gaussian-broadened golden-rule rate is then
\begin{equation}
k_{\mathrm{IC}}^{(\sigma)}=
\frac{2\pi|v_{\mathrm{IC}}|^2}{\hbar}
\int dE\,S_{\mathrm{IC}}(E)
\frac{\exp[-(E_e-E)^2/(2\sigma^2)]}{\sqrt{2\pi}\sigma}.
\label{eq:ic_rate}
\end{equation}
We use $\sigma=0.008$~Ha, the Gaussian width employed for the
two-mode calculations of ref~\citenum{Wang2023Nonradiative}.
Equivalently, this multiplies
$C(t)=e^{iE_et/\hbar}\langle\chi|e^{-i\hat H_gt/\hbar}|\chi\rangle$
by $\exp[-\sigma^2t^2/(2\hbar^2)]$ before time integration.
The plotted ratio divides the aggregate rate by the independently
evaluated harmonic-monomer rate at the same broadening and
nonadiabatic coupling. The common coupling prefactor cancels;
there is no additional division by the number of molecules because
the aggregate contains one normalized excitation.

\begin{figure}[tbp]
\centering
\includegraphics[width=0.98\linewidth]{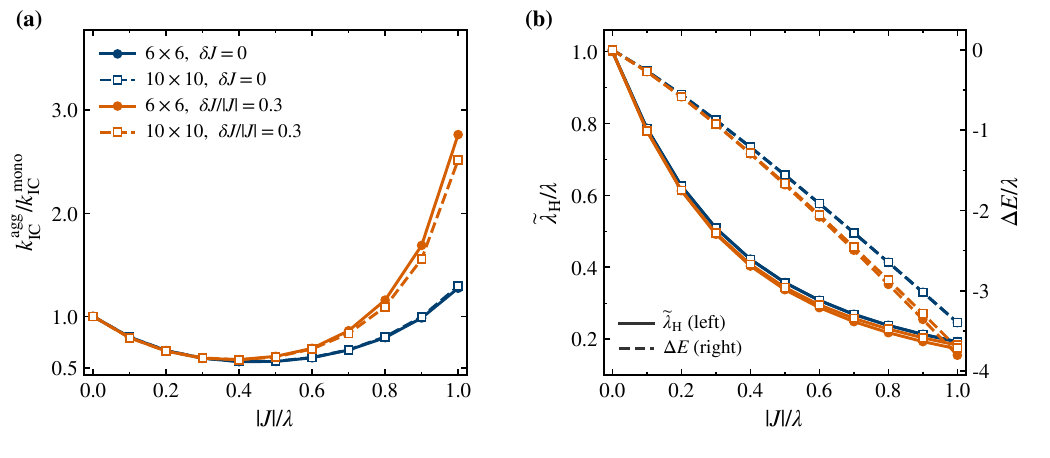}

\caption{Internal conversion and static vibronic observables on periodic $6\times6$ and $10\times10$ molecular lattices at zero temperature. Blue and orange denote $\delta J=0$ and $\delta J/|J|=0.3$, respectively; filled circles and open squares denote $6\times6$ and $10\times10$. (a) Aggregate-to-monomer IC rate ratios calculated with one-site TaChe, Burg reconstruction, and Gaussian broadening (\(\sigma=0.008~\mathrm{Ha}\)). Solid and dashed lines connect the \(6\times6\) and \(10\times10\) results, respectively. Both \(6\times6\) curves used a bond-dimension cap of \(D=64\) and 2048 Chebyshev moments. For the \(10\times10\) lattice with \(\delta J/|J|=0.3\), we retained 2048 moments throughout and increased the bond-dimension cap from \(D=72\) to \(D=96\) for \(x=|J|/\lambda>0.6\). For the \(10\times10\) lattice with \(\delta J=0\), we used \(D=72\). (b) Effective Holstein reorganization energy $\widetilde\lambda_H/\lambda$ (solid lines, left axis) and energy-gap shift $\Delta E/\lambda$ (dashed lines, right axis). }

\label{fig:ic}
\end{figure}

The rates in Fig.~\ref{fig:ic}(a) are reconstructed directly from
these scalar moments using Burg spectral maximum entropy, with a
relative covariance loading of $10^{-8}$, followed by the Gaussian
convolution in Eq.~\eqref{eq:ic_rate}. Panel (b) uses static observables evaluated
from the same initial TTNS states, independently of the spectral
reconstruction.

Both lattice sizes exhibit a nonmonotonic IC rate [Fig.~\ref{fig:ic}(a)]. The rate first decreases below the monomer value and reaches a minimum near $|J|/\lambda=0.4$. For the $6\times6$ lattice, the minimum ratios are $0.56$ without hopping fluctuations and $0.58$ with them. The two sizes give similar rates through this suppression regime, showing that the intermediate-coupling trend persists when the aggregate grows from $36$ to $100$ molecules.

At larger coupling, hopping fluctuations strongly increase the rate. At $|J|/\lambda=1$, the $6\times6$ ratios reach $1.27$ for $\delta J=0$ and $2.76$ for $\delta J/|J|=0.3$. The $10\times10$ aggregate remains close to the smaller lattice without hopping fluctuations, while its fluctuating-coupling rate rises to approximately $2.5$ times the monomer value.

To interpret these trends, we evaluate the accepting-mode distortion
field and its associated effective reorganization energy,
\begin{equation}
\mathcal D_H(\mathbf r)=\sum_i
\langle\Psi_e|\hat n_i\hat q_{i+\mathbf r,h}|\Psi_e\rangle,
\qquad
\widetilde\lambda_H=
\frac12\omega_h^2\sum_{\mathbf r}\mathcal D_H(\mathbf r)^2,
\label{eq:ic_vdf}
\end{equation}
where $\hat q_{ih}=\sqrt{\hbar/(2\omega_h)}
(\hat b_{ih}^{\dagger}+\hat b_{ih})$ is a mass-weighted coordinate.
We also calculate
$\Delta E=E_{e,\mathrm{agg}}-E_{e,\mathrm{mono}}$, using the
same ground-sector energy zero. The distortion-field definition
follows ref~\citenum{Wang2023Nonradiative}.
For the extended model, $\widetilde\lambda_H$ is the intramolecular
Holstein contribution, while $\Delta E$ includes the full Peierls
interaction.

Figure~\ref{fig:ic}(b) shows a rapid decrease in $\widetilde\lambda_H$ at weak coupling, followed by a more gradual reduction toward roughly one-fifth of the monomer value. The excitation gap continues to narrow as $|J|$ grows. The initial rate suppression is consistent with reduced local reorganization, while the strong-coupling increase accompanies the continuing gap reduction. Hopping fluctuations further lower the gap and reshape the vibronic transition spectrum, producing a much larger change in the rate than in the intramolecular distortion measure. 

\section{Conclusions and Outlooks}

In this work, we developed TaChe for computing quantum many-body spectra and dynamics in the one-site and two-site tangent spaces of TTNS. By expressing the projected Hamiltonian in terms of orthogonal tangent-space coefficients, TaChe eliminates the need for repeated state compression during the Chebyshev recurrence. The three-term recurrence further ensures memory efficiency, as only the two preceding Chebyshev vectors for the next vector are required, keeping memory usage independent of the expansion order. Combined with GPU-accelerated tensor contractions, this formulation enables efficient spectral calculations for the challenging molecular systems studied here. Through benchmarks spanning molecular vibronic spectra, thermal emission, charge transport, and internal-conversion, we demonstrate that TaChe provides an accurate and computationally efficient framework for accessing spectral and dynamical observables in strongly coupled electron--vibrational systems, including spatially extended molecular aggregates with up to $100$ molecules.

Looking forward, while we employ imaginary-time evolution to prepare thermal states in this work, alternative approaches \cite{Chen2018XTRG,Li2023tanTRG,Gao2026Born} may offer more efficient thermal-state preparation. The tangent-space formulation opens several promising directions for future development. First, the accuracy of the projected spectrum depends on both the reference state and the accessible tangent space. Exploring better reference states tailored to molecular excitations may provide an additional route to improving spectral accuracy in quantum chemistry. More importantly, because the tangent space forms a linear vector space, iterative calculations avoid repeated tensor truncation and can readily incorporate established numerical linear-algebra techniques, such as thick-restart Lanczos, reorthogonalization, and deflation, to improve convergence and spectral resolution. Extensions to dissipative molecular dynamics could also be built, with appropriate treatment of the complex spectrum. Beyond tree tensor networks, the single-quasiparticle ansatz for projected entangled-pair states represents excitations as coherent sums of local tensor replacements~\cite{VanderstraetenEtAl2015,VanderstraetenPEPS2019}. Extending the moment expansion to this tangent space would require explicit treatment of its nontrivial norm metric and approximate environment contractions. Taken together, these possibilities make the truncation-free tangent-space formulation a particularly promising framework for frequency-domain calculations of challenging quantum many-body systems.

\section{Data Availability}

The main source code for TaChe, as well as the data of this work, are publicly available at https://github.com/yangz15/Tangent-Chebyshev.
\begingroup\singlespacing
\begin{acknowledgement}
W.L. acknowledges support from the National Natural Science Foundation of China under Grant Nos. 22595405 and 22573088, the Guangdong Basic and Applied Basic Research Foundation under Grant No. 2026A1515010545, the Guangdong Provincial Quantum Science Strategic Initiative under Grant No. GDZX2503001, the Shenzhen Science and Technology Program under Grant Nos. KQTD20240729102028011 and JCYJ20250604140259001, and the Guangdong Basic Research Center of Excellence for Aggregate Science. Y. W. acknowledges support from the National Natural Science
Foundation of China under Grant No. 22325302. C.B.M. and J.v.D. acknowledge funding from the Munich Quantum Valley initiative, supported by the Bavarian state government with funds from the Hightech Agenda Bayern Plus. Z.S. acknowledges support from the National Natural Science Foundation of China under Grant Nos. T2350009 and 22433007, the Guangdong Basic and Applied Basic Research Foundation under Grant No. 2024A1515011185, the Shenzhen Science and Technology Program under Grant No. KQTD20240729102028011, and the Guangdong Basic Research Center of Excellence for Aggregate Science. The authors acknowledge the High Performance Computing Center at The Chinese University of Hong Kong, Shenzhen, for providing computational resources.
\end{acknowledgement}

\section*{Supporting Information}
The coCheb method, details of maximum entropy spectral reconstruction, and computational costs are introduced in the supporting information.
\bibliography{references}

\clearpage

\endgroup
\end{document}


\clearpage

\begingroup
\singlespacing
\tableofcontents
\endgroup
\clearpage
\section{The coCheb Method}
\label{sec:si-canonical}
\subsection{The canonical-orthogonalized TTNS Chebyshev method}

The canonical-orthogonalized TTNS Chebyshev method (coCheb) follows
the coCheMPS construction of Jiang et al.\cite{JiangEtAl2021}
Here the primitive states are binary TTNS. The method uses a
compressed Chebyshev sequence to construct a trial subspace and then
projects the Hermitian generator onto a canonically orthonormal basis
of that subspace. Let $|b\rangle$ be the response source and $\mathcal A$
the generator, for example $H-E_0$ for a ground-state spectrum or the
Hermitian Liouvillian for thermal emission. Choose $(c,a)$ with $a>0$
so that the spectral interval used for the primitive recurrence maps
inside $(-1,1)$. With $\mathcal C_D$ denoting
compression to the chosen bond dimensions and
$\widetilde{\mathcal A}=(\mathcal A-cI)/a$, their generation is
\begin{equation}
 \begin{aligned}
 |\chi_0\rangle&=\mathcal C_D[|b\rangle],\qquad
 |\chi_1\rangle=\mathcal C_D[\widetilde{\mathcal A}|\chi_0\rangle],\\
 |\chi_{n+1}\rangle&=\mathcal C_D[
 2\widetilde{\mathcal A}|\chi_n\rangle-|\chi_{n-1}\rangle].
 \end{aligned}
 \label{eq:si-cocheb-primitive}
\end{equation}
The compression in each step can be implemented by a variational fit
of a rank-capped TTNS to the entire recurrence target.
Collect $N$ primitive vectors into the column map
$Y_N=[|\chi_0\rangle,\ldots,|\chi_{N-1}\rangle]$ and form
\begin{equation}
 S=Y_N^{\dagger}Y_N,\qquad J=Y_N^{\dagger}\mathcal A Y_N,
 \qquad S=V\operatorname{diag}(\lambda_j)V^{\dagger}.
 \label{eq:si-cocheb-matrices}
\end{equation}
The matrix elements of $S$ and $J$ are contracted directly from the stored states, accounting for the independently compressed primitive vectors. Retain the eigenvalues
$\lambda_j>\tau_S\lambda_{\max}$, with $\tau_S>0$, and their
eigenvectors in $\Lambda_R$ and $V_R$. Canonical
orthogonalization\cite{Lowdin1970} gives
\begin{equation}
 \begin{aligned}
 X&=V_R\Lambda_R^{-1/2},\qquad Q=Y_N X,\\
 K_r&=X^{\dagger}JX,\qquad
 \boldsymbol\gamma=Q^{\dagger}|b\rangle.
 \end{aligned}
 \label{eq:si-canonical-basis}
\end{equation}
Then $Q^{\dagger}Q=I$, and $K_r=Q^{\dagger}\mathcal A Q$ is a
Hermitian effective generator. The basis $Q$ need only be represented
through $X$: if $f_i=\langle\chi_i|b\rangle$, then
$\boldsymbol\gamma=X^{\dagger}\boldsymbol f$.
Diagonalization, $K_r\boldsymbol y_j=\epsilon_j\boldsymbol y_j$,
provides the reduced spectral measure
\begin{equation}
 F_{\mathrm{co},N}(E)=\sum_j
 |\boldsymbol y_j^{\dagger}\boldsymbol\gamma|^2
 \delta(E-\epsilon_j).
 \label{eq:si-discrete-spectrum}
\end{equation}
The retained spectral weight is $\|Q^{\dagger}|b\rangle\|^2$,
and the source weight outside the reduced space is
$\|(I-QQ^{\dagger})|b\rangle\|^2$. Algorithm~\ref{alg:si-cocheb}
summarizes the construction for an autocorrelation spectrum.

\begin{algorithm}[tbp]
\caption{Canonical-orthogonalized Chebyshev method with binary TTNS}
\label{alg:si-cocheb}
\begin{algorithmic}[1]
\Require Hermitian generator $\mathcal A$, source $|b\rangle$, shift and scale $(c,a)$ with $a>0$, bond cap $D$, primitive count $N\geq1$, overlap cutoff $\tau_S$
\State $\widetilde{\mathcal A}\gets(\mathcal A-cI)/a$
\State $|\chi_0\rangle\gets\mathcal C_D[|b\rangle]$
\If{$N>1$}
  \State $|\chi_1\rangle\gets\mathcal C_D[\widetilde{\mathcal A}|\chi_0\rangle]$
\EndIf
\For{$n=1,\ldots,N-2$}
  \State $|\chi_{n+1}\rangle\gets\mathcal C_D[2\widetilde{\mathcal A}|\chi_n\rangle-|\chi_{n-1}\rangle]$
\EndFor
\State Contract $S_{ij}\gets\langle\chi_i|\chi_j\rangle$,
$J_{ij}\gets\langle\chi_i|\mathcal A|\chi_j\rangle$, and
$f_i\gets\langle\chi_i|b\rangle$
\State Diagonalize $S$ and retain $(\Lambda_R,V_R)$ with $\lambda_j>\tau_S\lambda_{\max}$
\If{the retained span is empty}
  \State \Return zero spectral weight
\EndIf
\State $X\gets V_R\Lambda_R^{-1/2}$,
$K_r\gets X^\dagger JX$,
$\boldsymbol\gamma\gets X^\dagger\boldsymbol f$
\State Solve $K_r\boldsymbol y_j=\epsilon_j\boldsymbol y_j$
\State \Return poles $\epsilon_j$ and weights $|\boldsymbol y_j^\dagger\boldsymbol\gamma|^2$
\end{algorithmic}
\end{algorithm}

\subsection{Spectral Readout in the Reduced Space}

The coCheb effective spectrum can be broadened directly from
Eq.~\eqref{eq:si-discrete-spectrum}, or reconstructed by a second
Chebyshev expansion.\cite{JiangEtAl2021} When using the first method, this method is equivalent to Krylov method if numerics are exact. With the second method, one may choose $c$ and $a$ to place the entire spectrum of $\widetilde K_r=(K_r-cI)/a$ inside $(-1,1)$.
For $x=(E-c)/a$, an $M$-term expansion gives
\begin{equation}
 \begin{aligned}
 F_{\mathrm{co},N,M}(E)&=\frac{g_0\widehat\mu_0+
 2\sum_{m=1}^{M-1}g_m\widehat\mu_mT_m(x)}
 {\pi a\sqrt{1-x^2}},\\
 \widehat\mu_m&=\boldsymbol\gamma^{\dagger}
 T_m(\widetilde K_r)\boldsymbol\gamma.
 \end{aligned}
 \label{eq:si-second-expansion}
\end{equation}
In this convention the Jackson coefficients are\cite{WeisseEtAl2006}
\begin{equation}
 g_m=\frac{(M-m+1)\cos(m\theta)+
 \sin(m\theta)\cot\theta}{M+1},\qquad
 \theta=\frac{\pi}{M+1}.
 \label{eq:si-jackson}
\end{equation}
Within the retained subspace, the second recurrence is ordinary matrix--vector arithmetic with $\widetilde K_r$ and requires no TTNS
compression. The primitive-space size $N$ controls the subspace approximation, and the second expansion order $M$ controls the
polynomial readout.

\section{Maximum-Entropy Reconstruction }
\label{sec:si-maxent}

\subsection{TaChe in a tangent Space}

Let $U$ be an isometric embedding of the chosen response space
$\mathscr V$, so that $U^{\dagger}U=I$ and $UU^{\dagger}=P$.
For a Hermitian generator $\mathcal A$ and physical source $|b\rangle$,
set $K=U^{\dagger}\mathcal A U$ and
$\boldsymbol b=U^{\dagger}|b\rangle$. The recurrence
\begin{equation}
 \begin{aligned}
 \boldsymbol t_0&=\boldsymbol b,\qquad
 \boldsymbol t_1=\widetilde K\boldsymbol b,\\
 \boldsymbol t_{n+1}&=2\widetilde K\boldsymbol t_n-
 \boldsymbol t_{n-1},\qquad
 \widetilde K=(K-cI)/a
 \end{aligned}
 \label{eq:si-chebyshev-recurrence}
\end{equation}
implies $\boldsymbol t_n=T_n(\widetilde K)\boldsymbol b$ by induction.
Both $U$ and $K$ remain fixed throughout the calculation. TaChe evaluates the moments $\mu_n=\boldsymbol b^{\dagger}\boldsymbol t_n$
directly in these orthonormal coordinates.

For self-correlations, the polynomial product formula yields
\begin{equation}
 \begin{aligned}
 \mu_{2n}&=2\boldsymbol t_n^{\dagger}\boldsymbol t_n-\mu_0,\\
 \mu_{2n+1}&=2\boldsymbol t_n^{\dagger}\boldsymbol t_{n+1}-\mu_1.
 \end{aligned}
 \label{eq:si-doubled-moments}
\end{equation}
These identities use a fixed Hermitian generator and the same source
on both sides. The benchmark calculations retain two scalar moments per
projected-generator application.

\subsection{From Chebyshev Moments to Angular Covariances}

For a self-correlation, the fixed-space spectral measure
$F_P(E)=\boldsymbol b^\dagger\delta(E-K)\boldsymbol b$ is nonnegative,
with integrated weight $\mu_0=\|\boldsymbol b\|^2$. We use the
undamped moments from Eq.~\eqref{eq:si-chebyshev-recurrence} or
Eq.~\eqref{eq:si-doubled-moments}. When $\mu_0=0$, the projected
source and its spectrum both vanish. Otherwise, let $p=M-1$ for
$M$ supplied moments and set $r_n=\mu_n/\mu_0$ for $0\leq n\leq p$.
The change of variables $E=c+a\cos\theta$ and the identity
$T_n(\cos\theta)=\cos(n\theta)$ give the constraints
\begin{equation}
 r_n=\frac{1}{2\pi}\int_{-\pi}^{\pi}
      P(\theta)\cos(n\theta)\,d\theta,\qquad 0\leq n\leq p .
 \label{eq:si-burg-moment-constraints}
\end{equation}
Here $P$ is the even angular density, expressed relative to
$d\theta/(2\pi)$. In particular, $r_0=1$. For a discrete input spectrum
the exact constraints are understood as integrals against a measure.
The real, even extension $r_{-n}=r_n$ defines a Toeplitz covariance
matrix $R_{ij}=r_{i-j}$ for $0\leq i,j\leq p$. Its quadratic form
is the average of a squared trigonometric polynomial against the
angular measure, so $R$ is positive semidefinite. It can be singular
when the projected spectral measure has only a few distinct lines.

\subsection{Burg Entropy and the Yule--Walker System}

Among positive even densities that satisfy
Eq.~\eqref{eq:si-burg-moment-constraints}, we maximize
Burg's spectral entropy rate, up to an additive constant and an
overall factor,\cite{Burg1975,ChoiCover1984}
\begin{equation}
 \mathscr S[P]=\frac{1}{2\pi}\int_{-\pi}^{\pi}
                \log P(\theta)\,d\theta .
 \label{eq:si-burg-objective}
\end{equation}
Variation with the moment constraints makes $1/P$ a real, even
trigonometric polynomial of degree at most $p$. For positive-definite
$R$, its spectral factor gives the unique maximum in the form
\begin{equation}
 P_{\mathrm{ME}}(\theta)=\frac{v_p}{|A_p(e^{-i\theta})|^2},
 \qquad A_p(z)=1+\sum_{j=1}^{p}\alpha_j z^j ,
 \label{eq:si-burg-density}
\end{equation}
where $A_p$ has no zeros in $|z|\leq1$. The coefficients follow
from the Yule--Walker equations
\begin{equation}
 \sum_{j=1}^{p} r_{|k-j|}\alpha_j=-r_k
 \quad (1\leq k\leq p),\qquad
 v_p=r_0+\sum_{j=1}^{p}\alpha_j r_j .
 \label{eq:si-yule-walker}
\end{equation}
Indeed, $A_p(e^{-i\theta})P_{\mathrm{ME}}(\theta)$ is
$v_p/\overline{A_p(e^{-i\theta})}$. Its integral against
$e^{ik\theta}$ vanishes for $k\geq1$ and equals $v_p$ for $k=0$,
giving Eq.~\eqref{eq:si-yule-walker}. The Toeplitz system can be
solved by Levinson recursion in $O(p^2)$ operations and $O(p)$
working storage.\cite{Levinson1946} When $p=0$, $P_{\mathrm{ME}}=1$.

A diagonal loading $R\mapsto R+\eta I$ with $\eta>0$ adds a
constant angular background to the covariance data. With the
moments normalized by $\mu_0$, this amounts to increasing $r_0$
from $1$ to $1+\eta$ before the solve and dividing the reconstructed
angular density by $1+\eta$ afterward. Equivalently, solve with
$r_0=1$ and replace $r_n$ by $r_n/(1+\eta)$ for $n>0$. This
preserves the integrated weight and perturbs the higher moment
constraints by a relative factor of order $\eta$. 
\subsection{Energy-Domain Readout}

Restoring the projected source weight and the Jacobian gives
\begin{equation}
 F_{\mathrm{ME}}(E)=
 \frac{\mu_0 P_{\mathrm{ME}}\!\left(\arccos x\right)}
      {\pi a\sqrt{1-x^2}},\qquad
 x=\frac{E-c}{a},\qquad |x|<1 ,
 \label{eq:si-burg-energy}
\end{equation}
with zero density outside the rescaling interval. This density is
nonnegative, integrates to $\mu_0$, and reproduces the supplied
Chebyshev moments when $\eta=0$, or the loaded moments when
$\eta>0$. For integrals and physical line broadening, a grid in
$\theta$ absorbs the endpoint Jacobian. For a normalized energy-domain
kernel $g$, the broadened spectrum is
\begin{equation}
 (F_{\mathrm{ME}}*g)(E)
 =\frac{\mu_0}{\pi}\int_0^\pi
 P_{\mathrm{ME}}(\theta)\,g(E-c-a\cos\theta)\,d\theta .
 \label{eq:si-burg-convolution}
\end{equation}
The input to this reconstruction is the scalar TaChe moment sequence.
Time correlations and the broadened conductivity resolvent can instead
be evaluated directly from the undamped moments. The use of
Chebyshev moments for spectral inference follows
Silver and R{\"o}der,\cite{SilverRoder1997} while
Eq.~\eqref{eq:si-burg-objective} specifies the entropy functional
used here.

\section{Computational and memory Cost}
\label{sec:si-cost}
\subsection{Computational Cost}
Let $L$ denote the number of physical leaves, $\chi$ denote the input and compressed output TTNS maximum bond
dimension, the Hamiltonian TTNO has maximum bond dimension $D$, and $d$ the maximum physical dimension of a leaf. For grouped
physical and auxiliary indices, $d$ denotes their product dimension. Our binary tree contains $L$ physical leaves and
$L-2$ auxiliary vertices. The auxiliary tensors have three virtual indices and physical dimension one. The following
estimates exclude reference preparation and use the maximum tensor dimensions.

The dominant operation in TaChe is the application of the projected Hamiltonian to a Chebyshev vector. In the one-site
space, contractions at an auxiliary vertex cost $O(D^2\chi^4+D^3\chi^2)$, while those at a physical leaf cost
$O(Dd\chi^2+Dd^2\chi)$. Contributions from different tangent insertions are accumulated recursively over the tree, so the
number of local contractions grows linearly with $L$. One projected Hamiltonian application therefore costs
\begin{equation}
 O\!\left(
 L[D^2\chi^4+D^3\chi^2+Dd\chi^2+Dd^2\chi]
 \right).
 \label{eq:si-cost-one}
\end{equation}
Gauge projection and frame transformations do not increase
these powers. When auxiliary-vertex contractions dominate and
$D\lesssim\chi^2$, the leading cost is $O(LD^2\chi^4)$.

For the enlarged two-site tangent space, a pair of trivalent vertices has four external virtual indices. Keeping the two
local TTNO tensors separate gives the conservative bound \begin{equation}
 O\!\left(
 L[d^2D^2\chi^5+d^3D^3\chi^4]
 \right)
 \label{eq:si-cost-two}
\end{equation}
per projected Hamiltonian application. This bound replaces
both endpoint physical dimensions by $d$; internal--internal
pairs have physical dimension one at both endpoints.
At fixed $d$ and $D$, the one-site and two-site actions thus scale as $O(L\chi^4)$ and
$O(L\chi^5)$, respectively.

One-site TDVP uses the same local Hamiltonian contractions.
Its cost per time step additionally depends on the number of
Hamiltonian applications required by the local exponential
solvers. 

For TTNO TTNS application in the regular Hilbert space, variational compression optimizes a TTNS with bond dimension
at most $\chi$ against the target $H|\psi\rangle$, without explicitly constructing the product. For $s$ one-site
fitting sweeps, its cost is
\begin{equation}
 O\!\left(
 sL[D^2\chi^4+D^3\chi^2+Dd\chi^2+Dd^2\chi]
 \right).
\end{equation}
The required number of sweeps depends on the initial guess and fitting tolerance.

Our implementation additionally performs a preliminary two-site fitting sweep. Its Hamiltonian contractions are
bounded by $O(L[d^2D^2\chi^5+d^3D^3\chi^4])$. On an internal--internal edge, splitting the updated tensor requires
an SVD of a matrix with dimensions at most $\chi^2\times\chi^2$, costing $O(\chi^6)$ before symmetry reductions.

\subsection{Memory Requirements}
\label{sec:si-storage}

A stored one-site response vector contains
$O(L[\chi^3+d\chi])$ coefficients. Including the two-site
blocks increases this to $O(L[\chi^4+d\chi^2])$. TaChe retains a fixed number of response vectors throughout
the recurrence. Consequently, storing $M$ moments requires
\begin{equation}
 \begin{aligned}
 &O\!\left(L[\chi^3+d\chi]+M\right),
 &&\text{one-site space},\\
 &O\!\left(L[\chi^4+d\chi^2]+M\right),
 &&\text{two-site space}.
 \end{aligned}
 \label{eq:si-memory-tache}
\end{equation}
Moment doubling yields two moments per projected Hamiltonian application.

In coCheb, retaining $K$ TTNS vectors together with their overlap and Hamiltonian matrices requires
$O(KL[\chi^3+d\chi]+K^2)$ storage. The final canonical orthogonalization and diagonalization add $O(K^3)$ work.
The tangent space Lanczos~\cite{KovalskaEtAl2026} likewise retains $K$ Krylov vectors.

\clearpage
\providecommand{\latin}[1]{#1}
\makeatletter
\providecommand{\doi}
  {\begingroup\let\do\@makeother\dospecials
  \catcode`\{=1 \catcode`\}=2 \doi@aux}
\providecommand{\doi@aux}[1]{\endgroup\texttt{#1}}
\makeatother
\providecommand*\mcitethebibliography{\thebibliography}
\csname @ifundefined\endcsname{endmcitethebibliography}
  {\let\endmcitethebibliography\endthebibliography}{}